\documentclass[prb,aps,twocolumn,showpacs,superscriptaddress,floatfix]{revtex4}
\usepackage{epsfig}
\usepackage{amsmath}
\usepackage{amssymb}
\usepackage{amsfonts}
\usepackage{mathptmx}
\usepackage{eucal}
\usepackage{bm}
\usepackage{epstopdf}

\DeclareMathOperator{\sgn}{\mathop{\mathrm{sgn}}}

\begin{document}

\title{Josephson $\varphi$-junctions based on structures with complex
normal/ferromagnet bilayer}
\author{S.~V.~Bakurskiy}
\affiliation{Faculty of Physics, M.V. Lomonosov Moscow State University, 119992 Leninskie
Gory, Moscow, Russia}
\affiliation{Skobeltsyn Institute of Nuclear Physics, Lomonosov Moscow State University
1(2), Leninskie gory, Moscow 119991, Russian Federation}
\author{N.~V.~Klenov}
\affiliation{Faculty of Physics, M.V. Lomonosov Moscow State University, 119992 Leninskie
Gory, Moscow, Russia}
\author{T.~Yu.~Karminskaya}
\affiliation{Skobeltsyn Institute of Nuclear Physics, Lomonosov Moscow State University
1(2), Leninskie gory, Moscow 119991, Russian Federation}
\author{M.~Yu.~Kupriyanov}
\affiliation{Skobeltsyn Institute of Nuclear Physics, Lomonosov Moscow State University
1(2), Leninskie gory, Moscow 119991, Russian Federation}
\author{A.~A.~Golubov}
\affiliation{Faculty of Science and Technology and MESA+ Institute for Nanotechnology,
University of Twente, 7500 AE Enschede, The Netherlands}
\date{\today }

\begin{abstract}
We demonstrate that Josephson devices with nontrivial phase difference $%
0<\varphi_g <\pi $ in the ground state can be realized in structures
composed from longitudinally oriented normal metal (N) and ferromagnet (F)
films in the weak link region. Oscillatory coupling across F-layer makes the
first harmonic in the current-phase relation relatively small, while
coupling across N-layer provides negative sign of the second harmonic. To
derive quantitative criteria for a $\varphi$-junction, we have solved
two-dimensional boundary-value problem in the frame of Usadel equations for
overlap and ramp geometries of S-NF-S structures. Our numerical estimates
show that $\varphi $-junctions can be fabricated using up-to-date technology.
\end{abstract}

\pacs{74.45.+c, 74.50.+r, 74.78.Fk, 85.25.Cp}
\maketitle

%

\section{Introduction}

The relation between supercurrent $I_{S}$ across a Josephson junction and
phase difference $\varphi$ between the phases of the order parameters of
superconducting (S) banks is an important characteristic of a Josephson
structure \cite{Likharev,RevG}. In standard SIS structures with tunnel type
of conductivity of a weak link, the current-phase relation (CPR) has the
sinusoidal form $I_{s}(\varphi)=A\sin (\varphi )$. On the other hand, in SNS
or SINIS junctions with metallic type of conductivity the smaller the
temperature $T$ the larger the deviations from the $\sin (\varphi )$ form 
\cite{Likharev} and $I_{S}(\varphi )$ achieves its maximum at $\pi /2\leq $ $%
\varphi \leq \pi $. In SIS junctions the amplitude $B$ of second harmonic in
CPR, $B\sin (2\varphi ),$ is of the second order in transmission coefficient
of the tunnel barrier I and therefore is negligibly small for all $T$. In
SNS structures the second CPR harmonic is also small in the vicinity of
critical temperature $T_{C}$ of superconductors, where $A\sim (T_{C}-T)$. At
low temperatures $T\ll T_{C}$, the coefficients $A$ and $B$ have comparable
magnitudes, thus giving rise to qualitative modifications of CPR shape with
decrease of $T.$

It is important to note that in all types of junctions discussed above the
ground state is achieved at $\varphi =0$, since at $\varphi =\pi $ a
junction is at nonequilibrium state.

The situation changes in Josephson structures involving ferromagnets as weak
link materials. The possibility of the so-called \textquotedblleft $\pi $%
-state\textquotedblright\ in SFS Josephson junctions (characterized by the
negative sign of the critical current $I_{C}$) was predicted theoretically
and observed experimentally [\onlinecite{RevG}-\onlinecite{AnwarLR}].
Contrary to traditional Josephson structures, in SFS devices it is possible
to have the ground state $\varphi _{g}=\pi $ (so-called $\pi $-junctions),
while the $\varphi =0$ corresponds to an unstable situation. It was proven
experimentally \cite{Rogalla,Feofanov} that $\pi $-junctions can be used as
on-chip $\pi $-phase shifters or $\pi $-batteries for self-biasing various
electronic quantum and classical circuits. It was proposed to use self $\pi $%
-biasing to decouple quantum circuits from environment or to replace
conventional inductance and strongly reduce the size of an elementary cell 
\cite{Ustinov}.

In some classical and quantum Josephson circuits it is even more interesting
to create on-chip $\varphi $-batteries. They are $\varphi $-junctions, the
structures having phase difference $\varphi _{g}=\varphi $, $(0<|\varphi
|<\pi )$ between superconducting electrodes in the ground state. The $%
\varphi $-states were first predicted by Mints\cite{Mints} for the case of
randomly distributed alternating $0-$ and $\pi -$ Josephson junctions along
grain boundaries in high $T_c$ cuprates with d-wave order parameter
symmetry. It was shown later that $\varphi $-junctions can be also realized
in the periodic array of $0$ and $\pi \ $SFS junctions \cite{Buzdin,Pugach}.
It was demonstrated that depending on the length of $0$ or $\pi $ segments
in the array, a modulated state with the average phase difference $\varphi
_{g}$ can be generated if the mismatch length between the segments is small.
This $\varphi _{g}$ can take any value within the interval $-\pi \leq
\varphi _{g}\leq \pi $. Despite strong constraints on parameter spread of
individual segments estimated in \cite{SPIE}, remarkable progress was
recently achieved on realization of $\varphi $-junctions in such arrays \cite%
{Gold}. 

In general, in order to implement a $\varphi $-junction one has use a
Josephson junction having non-sinusoidal current-phase relation, which, at
least, can be described by a sum of two terms 
\begin{equation}
I_{S}(\varphi )=Asin(\varphi )+Bsin(2\varphi ).  \label{CPR}
\end{equation}%
Moreover, the following special relationship between the amplitudes of the
CPR harmonics, $A,$ and, $B,$ is needed for existence of equilibrium stable
state \cite{Gold2007,Klenov} 
\begin{equation}
|B|>\left\vert A\right\vert /2,B<0.  \label{SR1}
\end{equation}%
In conventional junctions, the magnitude of $A$ is larger than that of $B$
and the inequalities (\ref{SR1}) are difficult to fulfill. However, in SFS
junctions in the vicinity of $0$ to $\pi $ transition the amplitude of first
harmonic in CPR is close to zero, thus opening an opportunity for making a $%
\varphi -$ battery, if $B$ can be made negative. %
It is well-known that SFS junctions with metallic type of conductivity, as
well as SIFS structures \cite{Vasenko,Vasenko1} with high transparencies of
SF interfaces have complex decay length of superconducting correlations
induced into F-layer $\xi _{H}=\xi _{1}+i\xi _{2}$. Unfortunately, the
conditions (\ref{SR1}) are violated in these types junctions since the $%
A\sim \exp \{-L/\xi _{1}\}\cos (L/\xi _{2})\ $, $B\sim -\exp \{-2L/\xi
_{1}\}\cos (2L/\xi _{2})$, and for $L=(\pi /2)\xi _{2}\ $ corresponding to
the first $0$-$\pi \ $ transition the second harmonic amplitude $B$ is
positive.

Quantitative calculations made in the framework of microscopic theory\cite%
{Buzdin1,Vinokur} confirm the above qualitative analysis. In Ref. \cite%
{Buzdin1,Vinokur} it was demonstrated that in SFS sandwiches with either
clean or dirty ferromagnetic metal interlayer the transition from $0$ to $%
\pi \ $ state is of the first order, that is $B>0 $ at any transition point%
\cite{RevB}.

It was suggested recently in \cite{Karminskaya}$^{-}$\cite{Bakurskiy1} to
fabricate the "current in plane" SFS devices having the weak link region
consisting from NF\ or FNF multilayers with the supercurrent flowing
parallel to FN interfaces. In these structures, superconductivity is induced
from the S banks into the normal (N) film, while F films serves as a source
of spin polarized electrons, which diffuse from F to N layer thus providing
an effective exchange field in a weak link. Its strength it can be controlled%
\cite{Bergeret,Fominov} by transparencies of NF interfaces, as well as by
the products of densities of states at the Fermi level, $N_{F},$ $N_{N},$
and film thicknesses, $d_{F},$ $d_{N}$ . It was shown in \cite{Karminskaya}$%
^{-}$\cite{Karminskaya4} that the reduction of effective exchange energy in
a weak link permits to increase the decay length from the scale of the order
of $\sim 1$ nm up to $\sim 100$ nm. The calculations performed in these
papers did not go beyond linear approximation in which the amplitude of the
second harmonic in the CPR is small. Therefore, the question of the
feasibility of $\varphi -$contacts in these structures has not been studied
and remains open to date. 

\begin{figure}[t]
\includegraphics[width=5cm]{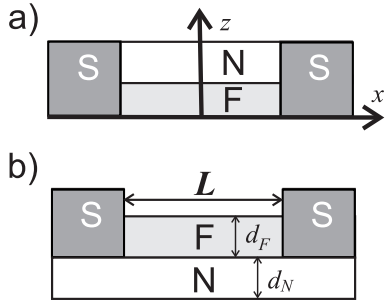} 
\caption{a) The $S-NF-S$ junction, b) the $SN-FN-NS$ junction. }
\label{types}
\end{figure}

The purpose of this paper is to demonstrate that the same "current in plane"
devices (see Fig. \ref{types}) can be used as effective $\varphi $-shifters.
The structure of the paper is the following. In Sec.\ref{CPR sec} we present
general qualitative discussion of the microscopic mechanisms leading to
formation of higher harmonics in the CPR. In Sec.\ref{Model} we formulate
quantitative approach in terms of Usadel equations. In Sec \ref{Ramp} the
criteria of $\varphi $-state existence are derived for ramp-type S-FN-S
structure. Section \ref{RTO_sec} shows the advantage of the other geometries
in order to realize $\varphi $-state. Finally in Sec.\ref{discuss} we
consider properties of real materials and estimate the possibility to
realize $\varphi $-states using up-to-date technology.

\section{CPR formation mechanisms \label{CPR sec}}

\begin{figure}[t]
\includegraphics[width=5cm]{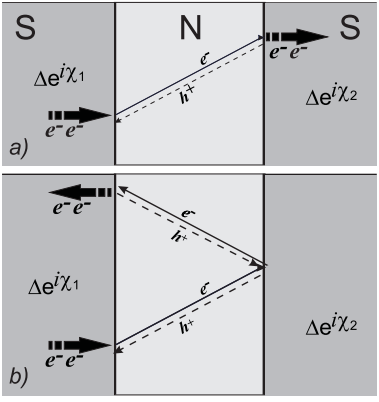} \vspace{-1mm}
\caption{Diagrams of the processes forming the first (a) and second~(b)
harmonics of the CPR in the SNS and SFS structures. }
\label{Andr}
\end{figure}

In this section we shall discuss microscopic processes which contribute to
formation of CPR in Josephson junctions. The physical reason leading to the
sign reversal of the coefficient $B$ in SFS junctions compared to that in
SNS structures can be understood from simple diagram shown in Fig.\ref{Andr}
illustrating the mechanisms of supercurrent transfer in double barrier
Josephson junctions.

Consider electron-like quasiparticle $e^{-}$ propagating across SINIS
structure towards the right electrode. This quasiparticle can be reflected
either in the Andreev or in the normal channel.

The result of the first process (see Fig.\ref{Andr}a) is generation in the
weak link region (with an amplitude proportional to $\exp (i\chi _{2})$) of
the hole $h^{+}$ propagating in the opposite direction. Andreev reflection
of this hole at the second interface (with an amplitude proportional to $%
\exp (-i\chi _{1})$) results in transfer of a Cooper pair from the left to
the right electrode with the rate proportional to the net coefficient of
Andreev reflection processes\cite{Furusaki, Furusaki1} at both SN
interfaces, $AR(\varphi )=\alpha (\varphi )\exp (i\varphi ),\varphi =(\chi
_{2}-\chi _{1}).$ The amplitude, $\alpha (\varphi ),$ depends on geometry of
a structure and on material parameters. Note that for given values of these
parameters $\alpha(\varphi )=\alpha(-\varphi ),$ according to the detailed
balance relations\cite{Furusaki}. Similar considerations show that a
quasiparticle $e^{-}$ moving towards the left electrode generates a Cooper
pair propagating from the right to the left interface with the rate
proportional to $AR(-\varphi )=\alpha (\varphi )\exp (-i\varphi ).$ The
difference between two processes described above determines a supercurrent $%
I_{S}$, which is proportional to $\sin (\varphi ).$

The result of the second process is the change (with an amplitude
proportional to $\exp (i\chi _{2})$) of the $e^{-}$ propagation direction to
the left electrode and nucleation of a Cooper pair 
and a hole propagating to the right electrode (with an amplitude
proportional to $\exp (-i\varphi )$). After normal reflection from the right
interface (with an amplitude proportional to $\exp (i\chi _{2})$) the hole
arrives at the left SN interface and closes this Andreev loop by generating
a Cooper pair in the left electrode and an electronic state (with an
amplitude proportional to $\exp (-i\chi _{1})).$ The Cooper pair have to
undergo a full reflection at SN interface, thus again a pair is generated
moving in the direction opposite to that in the main Andreev loop. The net
coefficient of this Andreev reflection process is $BR(\varphi )=\beta
(\varphi )\exp (2i\varphi ).$ For a quasiparticle $e^{-}$ moving in the weak
link towards the left electrode the same consideration leads to generation
of two Cooper pairs moving from the left to the right with the rate
proportional to $BR(-\varphi )=\beta (\varphi )\exp (-2i\varphi ).$ The
difference between these two processes determines a part of supercurrent $%
I_{S}$ proportional to $\sin (2\varphi ).$

We have shown that supercurrent components proportional to $\sin (\varphi )$
and $\sin (2\varphi )$ have opposite signs, and the coefficient $B $ in Eq.(%
\ref{CPR}) is negative. This statement is in a full agreement with
calculations of the CPR performed in the frame of microscopic theory of
superconductivity\cite{Likharev, RevG}. It is valid if a supercurrent across
a junction does not suppress superconductivity in S electrodes in the
vicinity of SN interfaces\cite{Ivanov,Zubkov,KuperD}. In addition, an
effective path of the particles in the second process discussed above is two
times larger than in the first one. This leads to stronger decay of the
second harmonic amplitude $B$ with increasing the distance $L$.

In SFS junctions the situation becomes more complicated. The exchange field, 
$H,$ in the weak link removes the spin degeneracy of quasiparticles. As a
result, one has to consider four types of Andreev's loops instead of two
loops discussed above. One should also take into account the fact that wave
function of a quasiparticle propagating through the weak link acquires an
additional phase shift $\varphi _{H}$ proportional to the magnitude of the
exchange field\cite{Beasley}. The sign of $\varphi _{H}$ depends on mutual
orientations between magnetization of the ferromagnetic film and the spin of
a quasiparticle. 
Taking into account these phase shifts and repeating arguments similar to
given above, one can show that the coefficients A and B in Eq.(\ref{CPR})
acquire additional factors $\cos (2\varphi _{H})$ and $\cos (4\varphi _{H}),$
respectively. At the point of $"0"$ - $"\pi "$ transition the coefficient $%
A=0$, that is $\varphi _{H}=\pi /4$. As a result, $\cos (4\varphi _{H})$
provides an additional factor, which changes the sign of the second harmonic
amplitude $B$ in SFS structures from negative to positive.

In the present study we will show that contrary to SFS devices with standard
geometry, it's possible to realize $\varphi$-junctions in the structures
shown in Fig. \ref{types}. Qualitatively, these structures are
superpositions of parallel SNS and SFS-channels, where supercurrent $%
I_{S}(\varphi)$ can be decomposed into two parts, $I_{N}(\varphi)$ and $%
I_{F}(\varphi)$, flowing across N and F films, respectively. For $L \ll
\xi_{N}$ and at sufficiently low temperatures $I_{N}(\varphi)$ has large
negative second CPR harmonic $B_{N}$. For $L > \xi_{1}$ supercurrent in the
SFS-channel exhibits damped oscillations as a function of $L$. In this
regime the second harmonic of CPR is negligibly small compared to the first
one. Large difference between decay lengths of superconducting correlations
in N and F-materials allows one to enter the regime when $\xi _{1} < L < \xi
_{N}$. In this case the first CPR harmonic $A=A_{N}+A_{F}$ can be made small
enough due to negative sign of $A_{F},$ while the second CPR harmonic $B
\approx B_{N}$ is negative, thus making it possible to fulfill the condition
(\ref{SR1}). Note that we are considering here the regime of finite
interface transparencies, when higher order harmonics decay fast with the
harmonic order. Therefore, it is sufficient to consider only the first and
the second harmonics of the CPR in all our subsequent discussions.

We show below that the mechanism described above indeed works in the
considered S-FN-S junctions, and we estimate corresponding parameter range
when $\varphi-$states can be realized.

\section{Model}

\label{Model}

We consider two types of symmetric multilayered structures shown
schematically on Fig.\ref{types}. The structures consist of a
superconducting (S) electrode contacting either the end-wall of a FN bilayer
(ramp type junctions) or the surface of F or N films (overlap junction
geometry). The FN bilayer consists of ferromagnetic (F) film and normal
metal (N) having a thickness $d_{F}$, and $d_{N}$ respectively. We suppose
that the conditions of a dirty limit are fulfilled for all metals and that
effective electron-phonon coupling constant is zero in F and N films. For
simplicity we assume that the parameters $\gamma _{BN}$ and $\gamma _{BF}$
which characterize the transparencies of NS and FS interfaces are large
enough%
\begin{equation}
\begin{array}{c}
\gamma _{BN}=\frac{R_{BN}\mathcal{A}_{BN}}{\rho _{N}\xi _{N}}\gg \frac{\rho
_{S}\xi _{S}}{\rho _{N}\xi _{N}} , \\ 
\gamma _{BF}=\frac{R_{BF}\mathcal{A}_{BF}}{\rho _{F}\xi _{F}}\gg \frac{\rho
_{S}\xi _{S}}{\rho _{F}\xi _{F}} ,%
\end{array}
\label{condgamma}
\end{equation}%
in order to neglect suppression of superconductivity in S parts of the
junctions. Here $R_{BN},R_{BF}$ and $\mathcal{A}_{BN},\mathcal{A}_{BF}$ are
the resistances and areas of the SN and SF interfaces, $\xi _{S},$ $\xi _{N}$
and $\xi _{F}$ are the decay lengths of S, N, F materials and $\rho _{S},$ $%
\rho _{N}$ and $\rho _{F}$ are their resistivities.

Under the above conditions the problem of calculation of the supercurrent in
the structures reduces to solution of the set of Usadel equations\cite{RevB,
RevV, Usadel} 
\begin{equation}
\frac{\xi ^{2}}{G_{\omega }}\partial \left[ G_{\omega }^{2}\partial \Phi
_{\omega }\right] -\frac{\widetilde{\omega }}{\pi T_{C}}\Phi _{\omega
}=0,G_{\omega }=\frac{\widetilde{\omega }}{\sqrt{\widetilde{\omega }%
^{2}+\Phi _{\omega }\Phi _{-\omega }^{\ast }}},  \label{UsadelFull}
\end{equation}%
where $\Phi _{\omega }$ and $G_{\omega }$ are Usadel Green's functions in $%
\Phi $ parametrization. They are $\Phi _{\omega ,N}$ and $G_{\omega ,N}$ or $%
\Phi _{\omega ,F}$ and $G_{\omega ,F}$ in N and F films correspondingly, $%
\omega =\pi T(2m+1)$ are Matsubara frequencies (m=0,1,2,...), $\widetilde{%
\omega }=\omega +iH,$ $H,$ is exchange field of ferromagnetic material, $\xi
^{2}=\xi _{N,F}^{2}=D_{N,F}/2\pi T_{C}$ for N and F layers respectively, $%
D_{N,F}$ are diffusion coefficients, $\partial =(\partial /\partial
x,\partial /\partial z)$ is 2D gradient operator. 
To write equations (\ref{UsadelFull}), we have chosen the $z$ and $x$ axis
in the directions, respectively, perpendicular and parallel to the plane of
N film and we have set the origin in the middle of structure at the free
interface of F-film (see Fig.\ref{types}).

The supercurrent $I_{S}(\varphi )$ can be calculated by integrating the
standard expressions for the current density $j_{N,F}(\varphi ,z)$ over the
junction cross-section: 
\begin{equation}
\begin{array}{c}
\frac{2ej_{N,F}(\varphi ,z)}{\pi T}=\sum\limits_{\omega =-\infty }^{\infty }%
\frac{iG_{\omega }^{2}}{\rho _{N,F}\widetilde{\omega }_{N,F}^{2}}\left[ \Phi
_{\omega }\frac{\partial \Phi _{-\omega }^{\ast }}{\partial x}-\Phi
_{-\omega }^{\ast }\frac{\partial \Phi _{\omega }}{\partial x}\right] , \\ 
I_{S}(\varphi )=W \int\limits_{0}^{d_{F}}j_{F}(\varphi
,z)dz+W\int\limits_{d_{F}}^{d_{F}+d_{N}}j_{N}(\varphi ,z)Wdz,%
\end{array}
\label{Ic_gen}
\end{equation}%
where $W$ is the width of the junctions, which is supposed to be small
compared to Josephson penetration depth. It is convenient to perform the
integration in (\ref{Ic_gen}) in F and N layers separately along the line
located at $x=0$, where $z$-component of supercurrent density vanishes by
symmetry.

Eq.(\ref{UsadelFull}) must be supplemented by the boundary conditions \cite%
{KL}. Since these conditions link the Usadel Green's functions corresponding
to the same Matsubara frequency $\omega $, we may simplify the notations by
omitting the subscript $\omega $. At the NF interface the boundary
conditions have the form:%
\begin{equation}
\begin{array}{c}
\gamma _{BFN}\xi _{F}\frac{\partial \Phi _{F}}{\partial z}=-\frac{G_{N}}{%
G_{F}}\left( \Phi _{F}-\frac{\widetilde{\omega }}{\omega }\Phi _{N}\right) ,
\\ 
\gamma _{BNF}\xi _{N}\frac{\partial \Phi _{N}}{\partial z}=\frac{G_{F}}{G_{N}%
}\left( \Phi _{N}-\frac{\omega }{\widetilde{\omega }}\Phi _{F}\right) ,%
\end{array}
\label{Gr6}
\end{equation}%
\begin{equation*}
\gamma _{BFN}=\frac{R_{BFN}\mathcal{A}_{BFN}}{\rho _{F}\xi _{F}}=\gamma
_{BNF}\frac{\rho _{F}\xi _{F}}{\rho _{N}\xi _{N}},
\end{equation*}%
where $R_{BFN}$ and $\mathcal{A}_{BFN}$ are the resistance and area of the
NF interface.

The conditions at free interfaces are 
\begin{equation}
\frac{\partial \Phi _{N}}{\partial n}=0,\;\frac{\partial \Phi _{F}}{\partial
n}=0.  \label{Gr5}
\end{equation}%
The partial derivatives in (\ref{Gr5}) are taken in the direction normal to
the boundary, so that $n$ can be either $z$ or $x$ depending on the
particular geometry of the structure.

In writing the boundary conditions at the interface with a superconductor,
we must take into account the fact that in our model we have ignored the
suppression of superconductivity in electrodes, so that in superconductor 
\begin{equation}
\Phi _{S}(\pm L/2)=\Delta \exp (\pm i\varphi /2),\;G_{S}=\frac{\omega }{%
\sqrt{\omega ^{2}+\Delta ^{2}}},  \label{FiS}
\end{equation}%
where $\Delta $ is magnitude of the order parameter in S banks. Therefore
for NS and FS interfaces we may write: 
\begin{subequations}
\begin{align}
\gamma _{BN}\xi _{N}\frac{\partial \Phi _{N}}{\partial n}& =\frac{G_{S}}{%
G_{N}}\left( \Phi _{N}-\Phi _{S}(\pm L/2)\right) ,  \label{Gr8} \\
\gamma _{BF}\xi _{F}\frac{\partial \Phi _{F}}{\partial n}& =\frac{G_{S}}{%
G_{F}}\left( \Phi _{F}-\frac{\widetilde{\omega }}{\omega }\Phi _{S}(\pm
L/2)\right) .  \label{Gr9}
\end{align}%
As in Eq. (\ref{Gr5}), $n$ in Eqs. (\ref{Gr8}), (\ref{Gr9}) is a normal
vector directed into material marked at derivative.

For the structure presented in Fig.\ref{types}a, the boundary-value problem (%
\ref{UsadelFull}) - (\ref{Gr9}) was solved analytically in the linear
approximation\cite{Karminskaya3, Karminskaya4}, i.e. under conditions 
\end{subequations}
\begin{equation}
G_{N}\equiv sgn(\omega ),\ G_{F}\equiv sgn(\omega ).  \label{Lin}
\end{equation}


In the present study we will go beyond linear approximation where
qualitatively new effects are found.


\section{Ramp-type geometry\label{Ramp}}

The ramp type Josephson junction has simplest geometry among the structures
shown in Fig.\ref{types}. It consists of the NF bilayer, laterally connected
with superconducting electrodes (see Fig.\ref{types}a).

In general, there are three characteristic decay lengths in the considered
structure\cite{Karminskaya}$^{,}$\cite{Karminskaya3}$^{,}$\cite{Crouzy}.
They are $\xi _{N},$ $\xi _{H}=\xi _{1}+i\xi _{2},$ and $\zeta =\zeta
_{1}+i\zeta _{2}.$ The first two lengths determine the decay and
oscillations of superconducting correlations far from FN interface, while
the last one describes their behavior in its vicinity. Similar length scale $%
\zeta $ occurs in a vicinity of a domain wall\cite{Chtchelkatchev}$^{-}$\cite%
{Crouzy}. In the latter, exchange field is averaged out for antiparallel
directions of magnetizations, and the decay length of superconducting
correlations becomes close to $\xi _{N}$. At FN interface, the flow of
spin-polarized electrons from F to N metal and reverse flow of unpolarized
electrons from N to F suppresses the exchange field in its vicinity to a
value smaller than that in a bulk ferromagnetic material thus providing the
existence of $\zeta $. Under certain set of parameters\cite{Karminskaya}
these lengths, $\zeta _{1}$, and, $\zeta _{2},$ can become comparable to $%
\xi _{N},$ which is typically much larger than $\xi _{1}$ and $\ \xi _{2}$,
which are equal to $\xi _{F}\sqrt{\pi T_{C}/H}$ for $H\gg \pi T_{C}$.

The existence of three decay lengths, $\xi _{N},$ $\zeta ,$ and $\xi _{H},$
should lead to appearance of three contributions to total supercurrent, $%
I_{N},$ $I_{FN}$ and $I_{F},$ respectively. The main contribution to $I_{N}$
component comes from a part of the supercurrent uniformly distributed in a
normal film. In accordance with the qualitative analysis carried out in
Section II, it is the only current component which provides a negative value
of the amplitude of the second harmonic $B$ in the current-phase relation.
The smaller the distance between electrodes $L$, the larger this
contribution. To realize a $\varphi -$contact, one must compensate for the
amplitude of the first harmonic, $A,$ in a total current to a value that
satisfies the requirement (\ref{SR1}). Contribution to $A$ from $I_{N}$ also
increases with decreasing $L$. Obviously, it's difficult to suppress the
coefficient $A $ due to the $I_{FN}$ contribution only, since $I_{FN}$ flows
through thin near-boundary layer. Therefore, strong reduction of $A$
required to satisfy the inequality (\ref{SR1}) can only be achieved as a
result of compensation of the currents $I_{N}$ and $I_{F}$ flowing in
opposite directions in N and F films far from FN interface. Note that the
oscillatory nature of the $I_{F}(L)$ dependence allows to satisfy
requirement (\ref{SR1}) in a certain range of $L$. The role of $I_{FN}$ in a
balance between $I_{N}$ and $I_{F}$ can be understood by solving the
boundary value problem (\ref{UsadelFull}) - (\ref{Gr9}) which admits an
analytic solution in some limiting cases.

\subsection{Limit of small $L.$\label{RampSmL}}

Solution of the boundary-value problem (\ref{UsadelFull})-(\ref{Gr9}) can be
simplified in the limit of small distance between superconducting electrodes 
\begin{equation}
L\ll \min \{\xi _{1},\xi _{N}\}.  \label{smallL1}
\end{equation}%
In this case one can neglect non-gradient terms in (\ref{UsadelFull}) and
obtain that contributions to the total current resulting from the
redistribution of currents near the FN interface cancel each other leading
to $I_{FN}=0$ (see Appendix \ref{A} for the details). As a result, the total
current $I_{S}(\varphi )$ is a sum of two terms only 
\begin{equation*}
I_{S}(\varphi )=I_{N}(\varphi )+I_{F}(\varphi ),
\end{equation*}%
\begin{equation}
\frac{2eI_{N}(\varphi )}{\pi TWd_{N}}=\frac{1}{\gamma _{BN}\xi _{N}\rho _{N}}%
\sum_{\omega =-\infty }^{\infty }\frac{\Delta ^{2}G_{N}G_{S}\sin (\varphi )}{%
\omega ^{2}},  \label{AIsmallL1}
\end{equation}%
\begin{equation}
\frac{2eI_{F}(\varphi )}{\pi TWd_{F}}=\frac{1}{\gamma _{BF}\xi _{F}\rho _{F}}%
\sum_{\omega =-\infty }^{\infty }\frac{\Delta ^{2}G_{N}G_{S}\sin (\varphi )}{%
\omega ^{2}},  \label{IF_small_1}
\end{equation}%
where $G_{N}=\frac{\omega }{\sqrt{\omega ^{2}+\Delta ^{2}\cos ^{2}(\frac{%
\varphi }{2})}}$. The currents $I_{N}(\varphi )$ and $I_{F}(\varphi )$ flow
independently across F and N parts of the weak link. The $I_{N,F}(\varphi )\ 
$ dependencies coincide with those calculated previously for double-barrier
junctions\cite{KL} in the case when $L$ lies within the interval defined by
the inequalities (\ref{smallL1}).

It follows from (\ref{AIsmallL1}), (\ref{IF_small_1}) that in the considered
limit neither the presence of a sharp FN boundary in the weak link region,
nor strong difference in transparencies of SN and SF interfaces lead to
intermixing of the supercurrents flowing in the F and N channels. It is also
seen that amplitude of the first harmonic of $I_{F}(\varphi )$ current
component is always positive and the requirement (\ref{SR1}) can not be
achieved.

\subsection{Limit of intermediate $L.$\label{RampIntL}}

For intermediate values of spacing between the S electrodes%
\begin{equation}
\xi _{1}\ll L\ll \xi _{N}  \label{interL1}
\end{equation}%
and for the values of suppression parameters at SN and SF interfaces
satisfying the conditions (\ref{condgamma}), the boundary problem (\ref%
{UsadelFull})-(\ref{Gr9}) can be solved analytically for sufficiently large
magnitude of suppression parameter $\gamma _{BFN}.$ It is shown in Appendix %
\ref{B} that under these restrictions in the first approximation we can
neglect the suppression of superconductivity in the N film due to proximity
with the F layer and find that 
\begin{equation}
\Phi _{N}=\Delta \cos (\frac{\varphi }{2})+i\frac{\Delta G_{S}\sin (\frac{%
\varphi }{2})}{\gamma _{BN}G_{N}}\frac{x}{\xi _{N}},~G_{N}=\frac{\omega }{%
\sqrt{\omega ^{2}+\Delta ^{2}\cos ^{2}(\frac{\varphi }{2})}},
\label{FN_in_N}
\end{equation}%
while spatial distribution of $\Phi _{F}(x,z)$ includes three terms: the
first two describe the influence of the N film, while the last one has the
form well known for SFS junctions \cite{RevG}$^{,}$\cite{RevB}$^{,}$\cite%
{RevV}.

Substitution of these solutions into expression for the supercurrent (\ref%
{Ic_gen}) leads to $I_{S}(\varphi )$ dependence consisting of three terms%
\begin{equation}
I_{S}(\varphi )=I_{N}(\varphi )+I_{F}(\varphi )+I_{FN}(\varphi ).
\label{IotFimiddle1}
\end{equation}%
Here $I_{N}(\varphi )$ is the supercurrent across the N layer. In the
considered approximation $I_{N}(\varphi )$ is given by the expression (\ref%
{AIsmallL1}). The second term in (\ref{IotFimiddle1}) equals to supercurrent
across SFS double barrier structure in the limit of small transparencies of
SF interfaces\cite{Golubov1}$^{,}$\cite{DBbuz}%
\begin{equation}
\frac{2eI_{F}(\varphi )}{\pi TWd_{F}}=\frac{\Delta ^{2}\sin \left( \varphi
\right) }{\gamma _{BF}^{2}\xi _{F}\rho _{F}}\sum_{\omega =-\infty }^{\infty }%
\frac{G_{S}^{2}}{\omega ^{2}\sqrt{\widetilde{\Omega }}\sinh \left(
2q_{L}\right) },  \label{IotFiSFS1}
\end{equation}%
where $q_{L}=L\sqrt{\widetilde{\Omega }}/2\xi _{F},$ $\widetilde{\Omega }%
=\left\vert \Omega \right\vert +iH\sgn(\Omega )/\pi T_{C},$ $%
\Omega=\omega/\pi T_{C}.$

The last contribution is shown in \ref{B} to contain three components 
\begin{equation}
I_{FN}(\varphi )=I_{FN1}(\varphi )+I_{FN2}(\varphi )+I_{FN3}(\varphi ).
\label{itemsINF}
\end{equation}%
with additional smallness parameters $\gamma _{BFN}^{-1}$ and $\gamma
_{BFN}^{-1}\xi_{F}/\xi _{N}$ compared to the current $I_{F}(\varphi )$ given
by Eq.(\ref{IotFiSFS1}). Nevertheless, these currents should be taken into
account in the analysis because they decay significantly slower than $%
I_{F}(\varphi )$ with increasing $L$.

\subsection{$\protect\varphi $-state existence \label{HSS_sec}}

The conditions for the implementation of a $\varphi -$contact are the
better, the larger the relative amplitude of the second harmonic which
increases at low temperatures. Therefore, low temperature regime is most
favorable for a $\varphi -$state. In the limit $T\ll T_{C}$ we can go from
summation to integration over $\omega$ in (\ref{AIsmallL1}), (\ref{IotFiSFS1}%
), (\ref{I1})- (\ref{I3}). From (\ref{AIsmallL1}) we have%
\begin{equation}
\frac{2eI_{N}(\varphi )}{Wd_{N}}=\frac{\Delta }{\gamma _{BN}\xi _{N}\rho _{N}%
}K(\sin \frac{\varphi }{2})\sin (\varphi ),  \label{IotFismallT}
\end{equation}%
where $K(x)$ is the complete elliptic integral of the first kind. Expanding
expression (\ref{IotFismallT}) in the Fourier series it is easy to obtain%
\begin{equation}
A_{N}=Q_{0}\frac{8}{\pi }\int\limits_{0}^{1}x^{2}\sqrt{1-x^{2}}%
K(x)dx=\Upsilon _{A}Q_{0},  \label{CPRA}
\end{equation}%
\begin{equation}
B_{N}=2A_{N}-\frac{32}{\pi }Q_{0}\int\limits_{0}^{1}x^{4}\sqrt{1-x^{2}}%
K(x)dx=\Upsilon _{B}Q_{0},  \label{CPRB}
\end{equation}%
where $Q_{0}=\Delta Wd_{N}/e\gamma _{BN}\xi _{N}\rho _{N},~A_{N},~B_{N}$ are
the first and the second harmonic amplitudes of $I_{N}(\varphi )$, 
\begin{eqnarray*}
\Upsilon _{A} &=&\frac{2\pi ^{2}}{\Gamma ^{2}(-\frac{1}{4})\Gamma ^{2}(\frac{%
7}{4})}{}\simeq 0.973, \\
\Upsilon _{B} &=&2\Upsilon _{A}-\frac{\pi }{2}{}_{3}F_{2}\left( \frac{1}{2},%
\frac{1}{2},\frac{5}{2};1,4;1\right) \simeq -0.146,
\end{eqnarray*}%
where $\Gamma (z)$ is Gamma-function and ${}_{p}F_{q}$ is generalized
hypergeometric function.

Evaluation of the sums in (\ref{IotFiSFS1}), (\ref{I1})- (\ref{I3}) can be
done for $H\gg \pi T_{C}$ and $T\ll T_{C}$ resulting in $I_{F}(\varphi
)=A_{F}\sin \left( \varphi \right) $ with%
\begin{equation}
A_{F}=P_{0}\frac{2}{\sqrt{h}}\exp \left( -\kappa L\right) \cos \left( \kappa
L+\frac{\pi }{4}\right),  \label{If_smallT}
\end{equation}%
$\kappa =\sqrt{h}/\sqrt{2}\xi _{F},$ $h=H/\pi T_{C}$ and $P_{0}=\Delta
Wd_{F}/e\gamma _{BF}^{2}\xi _{F}\rho _{F}.$ Substitution of (\ref{CPRA}), (%
\ref{CPRB}) into the inequalities (\ref{SR1}) gives $\varphi $-state
requirements for ramp-type structure 
\begin{equation}
\left\vert \Upsilon _{A}+\frac{1}{\varepsilon }\Psi (L)\right\vert
<2\left\vert \Upsilon _{B}\right\vert ,~\varepsilon =\frac{\sqrt{h}\gamma
_{BF}^{2}}{2\gamma _{BN}}\frac{d_{N}\xi _{F}\rho _{F}}{d_{F}\xi _{N}\rho _{N}%
},  \label{ineq2}
\end{equation}%
\begin{equation*}
\Psi (L)=\exp \left( -\kappa L\right) \cos \left( \kappa L+\frac{\pi }{4}%
\right) .
\end{equation*}%
This expression gives the limitation on geometrical and materials parameters
of the considered structures providing the existence of $\varphi $-junction.
Function $\Psi (L)$ has the first minimum at $\kappa L=\pi /2,$ $\Psi (\pi
/2\kappa )\approx -0.147$. For large values of $\varepsilon $ inequality (%
\ref{ineq2}) can not be fulfilled at any length $L$. Thus solutions exist
only in the area with upper limit 
\begin{equation}
\varepsilon <\frac{-\Psi (\pi /2\kappa )}{\Upsilon _{A}-2\left\vert \Upsilon
_{B}\right\vert }\approx 0.216.  \label{inEx}
\end{equation}%
At $\varepsilon \approx 0.216$ the left hand side of inequality (\ref{ineq2}%
) equals to its right hand part providing the nucleation of an interval of $%
\kappa L$ in which we can expect the formation of a $\varphi $-contact. This
interval increases with decrease of $\varepsilon $ and achieves its maximum
length 
\begin{equation}
1.00\lesssim \kappa L\lesssim 2.52,  \label{boun_kl}
\end{equation}%
at $\varepsilon =\frac{-\Psi (\pi /2\kappa )}{\Upsilon _{A}+2|\Upsilon _{B}|}%
\approx 0.116.$ It is necessary to note that at $\varepsilon =-\Psi (\pi
/2\kappa )/\Upsilon _{A}\approx 0.151$ 
%
there is a transformation of the left hand side local minimum in (\ref{ineq2}%
), which occurs at $\kappa L=\pi /2, $ into local maximum; so that at $%
\varepsilon \approx 0.116$ the both sides of (\ref{ineq2}) become equal to
each other, and the interval (\ref{boun_kl}) of $\varphi -$junction
existence subdivides into two parts. With a further decrease of $\varepsilon 
$ these parts are transformed into narrow bands, which are localized in the
vicinity of the $0-\pi $ transition point $(A_{N}+A_{F}=0)$; they take place
at $\kappa L=\pi /4$ and $\kappa L=5\pi /4.$ The width of the bands
decreases with decrease of $\varepsilon . $

Thus, our analysis has shown that for 
\begin{equation}
0.12\lesssim \varepsilon \lesssim 0.2  \label{int_e}
\end{equation}%
we can expect the formation of $\varphi -$junction in a sufficiently wide
range of distances $\Delta L$ between the electrodes determined by 
(\ref{ineq2}). Now we will take into the account the impact of the interface
term $I_{FN}(\varphi )$. In the considered approximations, it follows from (%
\ref{I1})- (\ref{I3}) that

\begin{equation}
\text{{\small $I_{FN1}(\varphi )=\frac{2U_{0}\xi _{F}\exp \left( -\frac{%
\kappa L}{2}\right) \cos \left( \frac{\kappa L}{2}-\frac{\pi }{4}\right) }{%
\gamma _{BF}\gamma _{BN}\xi _{N}h^{3/2}}$}}\sin \left( \varphi \right) ,
\label{I1s}
\end{equation}

\begin{equation}
\text{{\small $I_{FN2}(\varphi )=-$}}\frac{\sqrt{2}U_{0}\xi _{F}}{%
4h^{3/2}\gamma _{BN}\gamma _{BFN}\xi _{N}}\sin \left( \varphi \right) K(\sin 
\frac{\varphi }{2}),  \label{I2s}
\end{equation}

\begin{equation}
I_{FN3}(\varphi )=-\text{{\small $\frac{2U_{0}\exp \left( -\frac{\kappa L}{2}%
\right) \sin \left( \frac{\kappa L}{2}\right) }{h\gamma _{BF}}$}}\sin \left(
\varphi \right) K(\sin \frac{\varphi }{2}),  \label{I3s}
\end{equation}%
where $U_{0}=\Delta W/e\gamma _{BFN}\rho _{F}.$ In the range of distances
between the electrodes $\pi /4<\kappa L<5\pi /4$ currents $I_{FN2}(\varphi )$
and $I_{FN3}(\varphi )$ are negative. 
These contributions have the same form of CPR as it is for the $%
I_{N}(\varphi )$ term, and due to negative sign suppress the magnitude of
supercurrent across the junction thus making the inequality (\ref{ineq2})
easier to perform. The requirement $B<0$ imposes additional restriction on
the value of the suppression parameter $\gamma _{BFN}$ 
\begin{figure}[t]
\includegraphics[width=9cm]{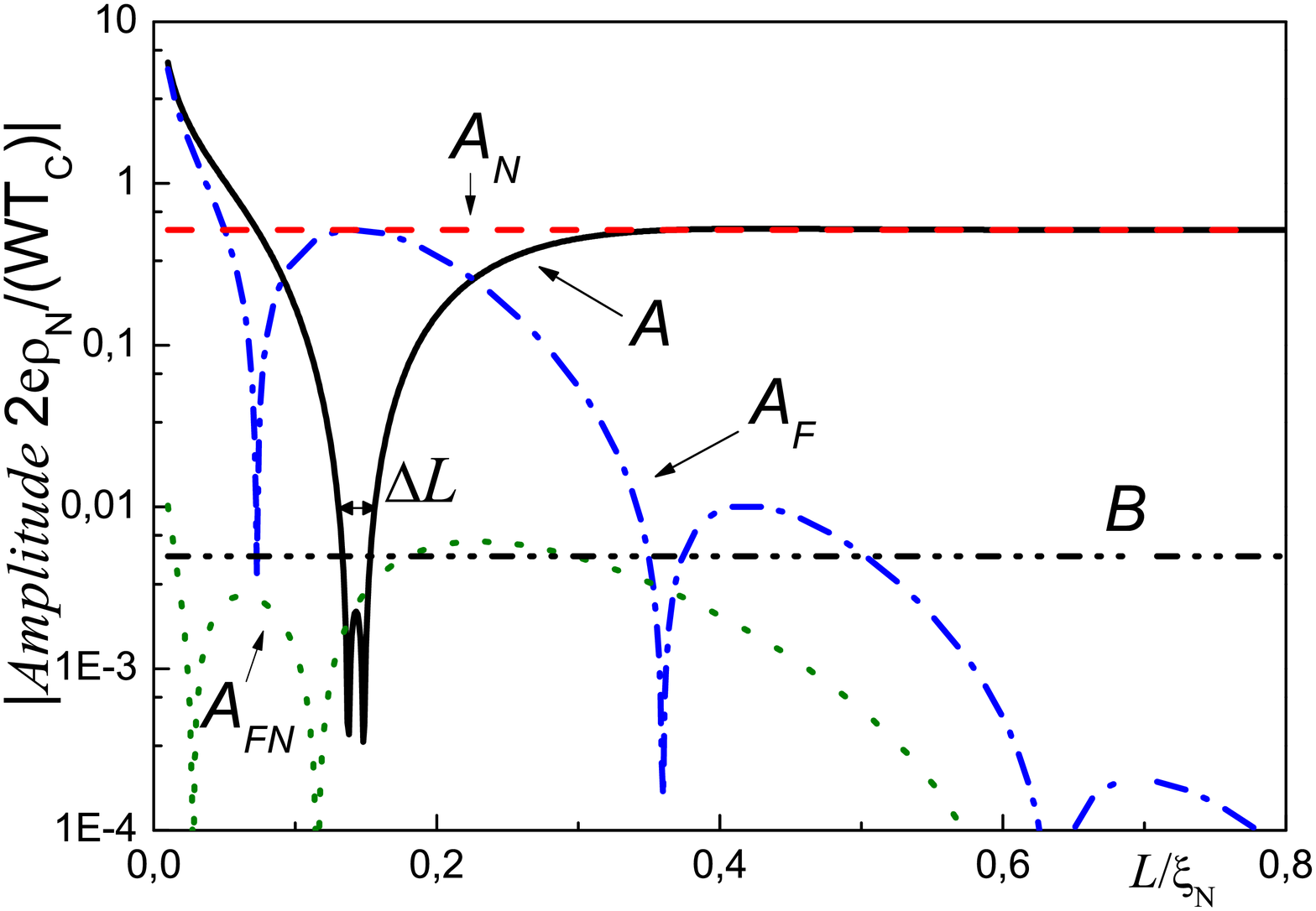} \vspace{-3mm}
\par
\vspace{-2mm}
\caption{Analytically derived amplitudes $A$ and $B$ in the CPR of ramp
S-NF-S structure ($d_{N}=0.1\protect\xi _{N},~d_{F}=0.65\protect\xi _{N}$)
and their components $A_{N},~A_{F},~A_{FN}$ versus electrode spacing $L$ at $%
T = 0.7 T_{C}$. Also enhanced interval of $\protect\varphi $-state, $\Delta
L,$ is marked.}
\label{analcurves}
\end{figure}
\begin{equation}
\gamma _{BFN}>\frac{\rho _{N}\xi _{N}}{hd_{N}\rho _{F}}\left( \frac{\xi _{F}%
}{\xi _{N}\gamma _{BFN}h^{1/2}}+\frac{\gamma _{BN}}{\gamma _{BF}}\right) .
\label{InEqGamBFN}
\end{equation}%
In derivation of this inequality we have used the fact that in the range of
distances between the electrodes $\pi /4<\kappa L<5\pi /4$ depending on $%
\kappa L$ \ factor in (\ref{I3s}) is of the order of unity. It follows from (%
\ref{InEqGamBFN}) that for a fixed value of $\gamma _{BFN}$ domain of $%
\varphi $-junction existence extends with increase of thickness of normal
films $d_{N}$ and this domain disappears if $d_{N}$ becomes smaller than the
critical value, $d_{NC},$ 
\begin{equation}
d_{NC}=\frac{\rho _{N}\xi _{N}}{h\rho _{F}\gamma _{BFN}}\left( \frac{\xi _{F}%
}{\xi _{N}\gamma _{BFN}h^{1/2}}+\frac{\gamma _{BN}}{\gamma _{BF}}\right) .
\label{IndN}
\end{equation}%
The existence of the critical thickness $d_{NC}$ follows from the fact that
the amplitude $B$ in $I_{N}$ is proportional to $d_{N},$ 
while in $I_{FN}$ term the parameter $B$ is independent on $d_{N}.$ The sign
of {\small $I_{FN1}(\varphi )$ }\ is positive for $\pi /4<\kappa L<3\pi /4$
and negative for $3\pi /4<\kappa L<5\pi /4$ thus providing an advantage for
a $\varphi -$junction realization for the lengths which belong to the second
interval.

Figure \ref{analcurves} illustrates our analysis. The solid line in Fig.\ref%
{analcurves} is the modulus of the amplitude of the first harmonic in CPR as
a function of distance $L$ between S electrodes. It is the result of
summation 
of the two contributions following from Eqs. (\ref{IotFiSFS1}) (dash-dotted
line) and (\ref{AIsmallL1}) (dashed line). The dash-dot-dotted line in Fig %
\ref{analcurves} is the amplitude of the second harmonic of the CPR
following from (\ref{AIsmallL1}). The dotted line is $I_{FN}(L)$ calculated
from (\ref{itemsINF}), (\ref{I1})- (\ref{I3}). All calculations have been
done for a set of parameters $d_{N}=0.1\xi _{N}$, $d_{F}=0.65\xi _{N}$, $%
\gamma _{BN}=0.1$, $\gamma _{BF}=1$, $\gamma _{BNF}=10$, $\xi _{F}=0.1\xi
_{N}$, $\rho _{N}=\rho _{F}$, $T=0.7T_{C}$, $H=10T_{C}$. These parameters
are close to those in real experimental situation. All the amplitudes were
normalized on factor $(2e\rho _{N}/(WT_{C}))^{-1}.$ It is evident that there
is an interval of $L$, for which the currents in N and F layers flow in
opposite directions. As a result of the addition of these currents the
points of $0-\pi $ transitions start to be closer to each other. It is seen
that in the entire region between these points, the inequality (\ref{SR1})
is fulfilled. This is exactly the $L-$interval, inside which a $\varphi -$%
junction can be realized. It is also seen that contribution of $I_{FN}$ part
into the full current is small and in accordance with our analisys does not
play a noticeable role. 
\begin{figure}[t]
\includegraphics[width=9cm]{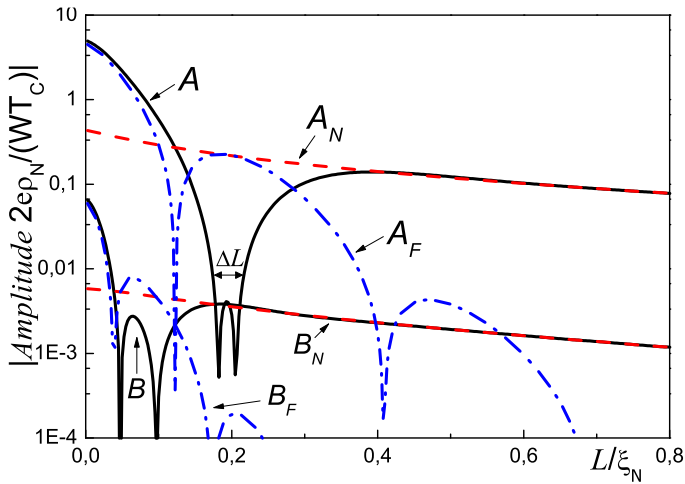} \vspace{-6mm}
\caption{Numerically calculated amplitudes $A$ and $B$ in the CPR of ramp
S-NF-S structure ($d_{N}=0.1\protect\xi _{N},~d_{F}=1.06\protect\xi _{N}$)
and their components $A_{N},~A_{F},~B_{N},~B_{F}$ versus electrode spacing $%
L $ at $T=0.7T_{C}$. In correspondence with Fig.\protect\ref{analcurves}
parameters are chosen to form enhanced $\protect\varphi $-state interval
marked by "$\Delta L$".}
\label{fi_HSS_2}
\end{figure}
\begin{figure}[bh]
\includegraphics[width=9cm]{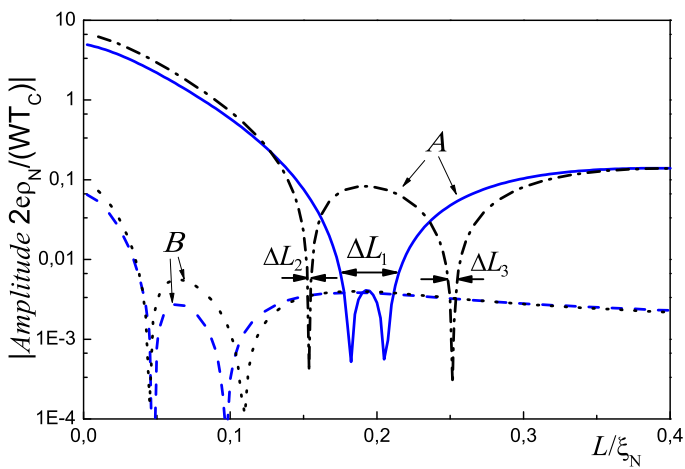} \vspace{-8mm}
\caption{Numerically calculated CPR amplitudes $A$ and $B$ versus electrode
spacing $L$ for S-FN-S structures with $d_{F}=1.06\protect\xi _{N}$ (solid
and dashed lines respectively) and $d_{F}=1.4\protect\xi _{N}$ (dash-dotted
and dotted lines). It is clear that enhanced $\protect\varphi $-interval $%
\Delta L_{1}$ formed in the first case is much larger than pair of ordinary $%
\protect\varphi $-intervals $\Delta L_{2}$ and $\Delta L_{3}$ in the second
one.}
\label{Windows}
\end{figure}
%

The boundary problem (\ref{UsadelFull})-(\ref{Gr9}) has been solved
numerically for the same set of junction parameters except $d_{F}.$ The
results of calculations for $d_{F}=1.06\xi _{N}$ and $d_{F}=1.4\xi _{N}$ are
shown in Fig.\ref{fi_HSS_2} and Fig.\ref{Windows}. The solid lines in Fig.%
\ref{fi_HSS_2} are the modulus of the amplitudes of the first, $A,$ and the
second, $B,$ harmonic of CPR as a function of distance $L$ between S
electrodes. The dashed and dash-dotted lines demonstrate the contributions
to these amplitudes from the currents flowing in N and F films,
respectively. All the amplitudes were normalized on the same factor $(2e\rho
_{N}/(WT_{C}))^{-1}.$ It is seen that the main difference between analytical
solutions presented in Fig.\ref{analcurves} and the curves calculated
numerically are located in region of small $L.$ It is also seen that
amplitudes of first and second harmonics of the part of the current flowing
in the N film slightly decay with $L$ increase. The points of $0-\pi $
transition of the first harmonic amplitude of the part of the current
flowing in the F layer is slightly shifted to the right, toward larger $L$.
It is also seen that the amplitude of the second harmonic, $B_{F},$ in the
interval of interest in the vicinity of $L\approx 0.2\xi _{N}$ is negligibly
small compared to the magnitude of, $B_{N}$. As a result, the shape of $A(L)$
curves in Fig.\ref{analcurves} and Fig.\ref{fi_HSS_2} is nearly the same,
with a little bit larger interval of $\varphi -$junction existence for the
curve calculated numerically.

Figure \ref{Windows} demonstrates the same $A(L)$ and $B(L)$ dependencies as
in Fig.\ref{fi_HSS_2} (solid and dashed lines) together with $A(L)$ and $%
B(L) $ curves calculated for $d_{F}=1.4\xi _{N}$ (dash-doted and dotted
lines). It is clearly seen that for larger $d_{F}$ we get out of the
interval (\ref{int_e}) and instead of relatively large zone $\Delta L_{1}$
may have $\varphi -$junction in two very narrow intervals $\Delta L_{2}$ and 
$\Delta L_{3}$ located in the vicinity of $0-\pi $ transitions of the first
harmonic amplitude $A.$

\section{Ramp type overlap (RTO) junctions \label{RTO_sec}}

\begin{figure}[t]
\includegraphics[width=6cm]{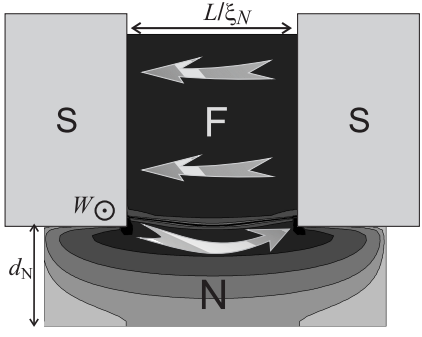} \vspace{-5mm} %
\caption{Current distribution along RTO-type SN-FN-NS structure at $L=0.63%
\protect\xi _{N}$, $d_{N}=\protect\xi _{N}$, $d_{F}=2\protect\xi _{N}$ and $%
T=0.7T_{C}$. The intensity of gray color shows current density in direction
indicated by arrows.}
\label{RTO1}
\end{figure}

Conditions for the existence of $\varphi -$junction (\ref{boun_kl}), (\ref%
{int_e}) can be improved by slight modifications of contact geometry,
namely, by using a combination of ramp and overlap configurations, as it is
shown in Fig.\ref{types}b. Fig.\ref{RTO1} demonstrates numerically
calculated spatial distribution of supercurrent in RTO $\varphi $-junction
at Josephson phase $\varphi =\pi /2$. The current density is presented by
darkness and the arrows give flows directions. The relative smallness of the
first harmonics amplitude is provided by opposite currents in N and F films.
The main feature of the ramp-overlap geometry is seen to be specific current
distribution in the normal layer leading to another CPR shape with
dependence on thickness $d_{N}$. Further, the current $I_{N}$ should
saturate as a function of $d_{N}$, 
%
since normal film regions located at distances larger than $\xi _{N}$ from
SN interface are practically excluded from the process of supercurrent
transfer due to exponential decay of proximity-induced superconducting
correlations \cite{KLO}. The specific geometry of the RTO structures makes
theoretical analysis of the processes more complex than in ramp contact.
Nevertheless, it is possible to find analytical expressions for supercurrent
in these structures and to show that the range of parameters providing the
existence of $\varphi -$state is broader than in the ramp type configuration.

To prove this statement, we consider the RTO structure in most practical
case of thin N film 
\begin{equation}
d_{N}\ll \xi _{N}  \label{condThinN1}
\end{equation}%
and sufficiently large $\gamma _{BFN}$ providing negligibly small
suppression of superconductivity in N film due to proximity with F layer. We
will assume additionally that electrode spacing $L$ is also small%
\begin{equation}
L\ll \xi _{N},  \label{smL1}
\end{equation}%
in order to have nonsinusoidal CPR. Under these conditions we can at the
first step consider the Josephson effect in overlap SN-N-NS structure. Then,
at the second step we will use the obtained solutions to calculate
supercurrent flowing across the F part of the RTO structure. The details of
calculations are summarized in Appendices \ref{C} and \ref{D}. They give
that the supercurrent%
\begin{equation}
I_{S}(\varphi )=I_{N}(\varphi )+I_{F}(\varphi )+I_{FN}(\varphi )
\label{IsRTO}
\end{equation}%
consists of three components. Expression for the part of current flowing
across N film has the form 
\begin{equation}
\frac{2eI_{N}(\varphi )}{\pi TWd_{N}}=\frac{2}{\rho _{N}\xi _{N}\sqrt{\gamma
_{BM}}}\sum_{\omega =-\infty }^{\infty }\frac{r^{2}\delta ^{2}\sin \varphi 
\sqrt{\left( \Omega \gamma _{BM}+G_{S}\right) }}{\sqrt{2\Omega \mu
^{2}\left( \sqrt{\Omega ^{2}+r^{2}\delta ^{2}}+\mu \right) }},
\label{RTOIN1}
\end{equation}%
where $r=G_{S}/\left( \Omega \gamma _{BM}+G_{S}\right) ,$ $\gamma
_{BM}=\gamma _{BN}d_{N}/\xi _{N}$ and $\mu =\sqrt{\Omega ^{2}+r^{2}\delta
^{2}\cos ^{2}(\varphi /2)},$ $\delta =\Delta /\pi T_{C}.$%

The $I_{F}(\varphi )$ term in (\ref{IsRTO}) is the current through one
dimensional double barrier SFS structure defined by Eq. (\ref{IotFiSFS1}),
while $I_{FN}(\varphi )$ is FN-interface term shown in \ref{D}. We provide
sufficient smallness and neglect it in the following estimations.

\begin{figure}[b]
\includegraphics[width=8cm]{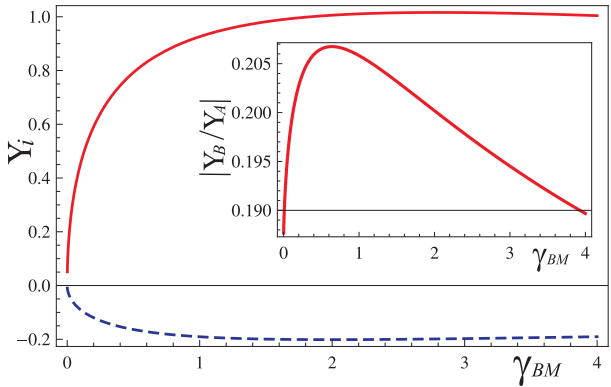} \vspace{-3mm}
\caption{The amplitudes of the first harmonic $\Upsilon _{A}$ (solid line)
and the second one $\Upsilon _{B}$ (dashed line) normalized on $2W\Delta/e%
\protect\rho_{N}\protect\gamma_{BN}$ versus reduced thickness $\protect%
\gamma _{BM}$. Inset shows the ratio of harmonics $|\Upsilon _{B}/\Upsilon
_{A}|$ versus $\protect\gamma _{BM} $.}
\label{RTO_harm}
\end{figure}
%
%
%
As we discussed above, the larger the relative amplitude of the second
harmonic (or the lower the temperature of a junction compare to $T_{C}$),
the better the conditions for the implementation of a $\varphi$-contact. In
the limit $T\ll T_{C}$ we can transform from summation to integration over $%
\omega $ in (\ref{RTOIN1}) and calculate numerically the dependence of
amplitudes $A$ and $B$ 
\begin{equation}
A_{N}=\frac{2W\Delta }{e\rho _{N}\gamma _{BN}}\Upsilon _{A},
\end{equation}%
\begin{equation}
B_{N}=\frac{2W\Delta }{e\rho _{N}\gamma _{BN}}\Upsilon _{B}
\end{equation}%
on suppression parameter $\gamma _{BM}.$ The calculated dependencies of
functions $\Upsilon _{A}(\gamma_{BM})$ and $|\Upsilon _{B}|(\gamma_{BM})$
are presented in Fig.\ref{RTO_harm}. It is seen that both $\Upsilon _{A}$
and $|\Upsilon _{B}|$ increase with increasing of $\gamma _{BM}$ and
saturate at $\gamma _{BM}\approx1.$ Inset in Fig.\ref{RTO_harm} shows the
ratio of the harmonics $|\Upsilon _{B}/\Upsilon _{A}|$ as a function of $%
\gamma _{BM}$. It achieves maximum at $\gamma _{BM}\approx 0.64,$ thus it
determines the optimal values of normalized amplitudes of the first $%
\Upsilon _{A}\approx 0.844$ and the second $\Upsilon _{B}\approx -0.175$
harmonics of the current flowing in the N layer. It is seen from the inset
in Fig.\ref{RTO_harm}, that the ratio $|\Upsilon _{B}/\Upsilon _{A}|$ is
slowly decreasing function of $\gamma_{BM}$. Therefore, the estimates given
below for $\gamma_{BM}=0.64$ are applicable in a wide parameter range $%
0.5\leq \gamma_{BM}\leq 10$.

Taking into account these values, we can write down the condition of $%
\varphi $-state existence similar to (\ref{ineq2})%
\begin{equation}
\left\vert \Upsilon _{A}+\frac{1}{\varepsilon }\Psi (L)\right\vert \leq
2\left\vert \Upsilon _{B}\right\vert ,~\varepsilon =\frac{\sqrt{h}\gamma
_{BF}^{2}}{\gamma _{BN}}\frac{\xi _{F}\rho _{F}}{d_{F}\rho _{N}},
\label{ineq3}
\end{equation}%
\begin{equation*}
\Psi (L)=\exp \left( -\kappa L\right) \cos \left( \kappa L+\frac{\pi }{4}%
\right) ,
\end{equation*}%
with slightly modified dimensionless parameter $\varepsilon $. The wide
region of $\varphi $-state still exists 
if $\varepsilon $ is within the interval 
\begin{equation}
0.123\lesssim \varepsilon \lesssim 0.298  \label{ineq4}
\end{equation}%
%
%
%
%
for $\kappa L$ that satisfies the condition (\ref{ineq3}). As follows from (%
\ref{ineq3}), interval of $\kappa L$ product gains its maximum length 
\begin{equation}
0.94\lesssim \kappa L\lesssim 2.72,  \label{boun_kl3}
\end{equation}%
at $\varepsilon=0.123$. It is seen that these intervals are slightly larger
than those given by (\ref{boun_kl}) for the ramp type geometry. 
\begin{figure}[t]
\includegraphics[width=8cm]{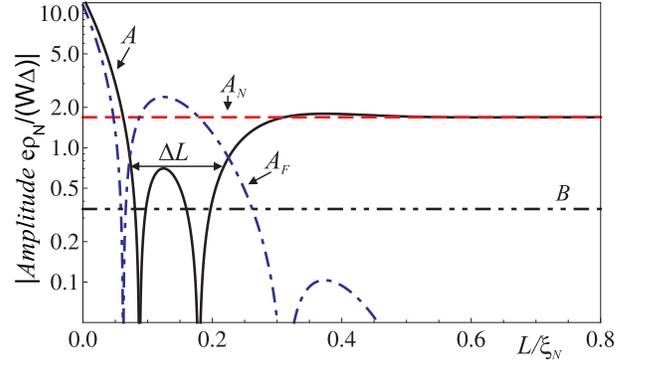} \vspace{-3mm}
\caption{{}The amplitudes of CPR harmonics $A,~A_{N},~A_{F,}~B$ versus
electrode spacing $L$ for RTO structure at $T\ll T_{C},$ $\protect\gamma %
_{BM}=0.64$ and $\protect\varepsilon =0.123$. The mark "$\Delta L$" shows
enhanced $\protect\varphi $-state interval.}
\label{RTO_ideal}
\end{figure}

Fig.\ref{RTO_ideal} shows the interval of $\varphi $-state existence, $%
\Delta L,$ in the ideal case of $T\ll T_{C},$ $\gamma _{BM}=0.64$ and $%
\varepsilon =0.123$. The corresponding set of parameters $d_{N}=0.64\xi _{N}$%
, $d_{F}=1.45\xi _{N}$, $\gamma _{BN}=1$, $\gamma _{BF}=1$, $\xi _{F}=0.1\xi
_{N}$, $\rho _{N}=\rho _{F}$, $H=10T_{C}$ was substituted in (\ref{IotFiSFS1}%
), (\ref{RTOIN1}). The solid line is a modulus of the first harmonic
amplitude, $A$, its normal, $A_{N},$ and ferromagnetic, $A_{F},$ parts are
presented by dashed and dash-dotted lines respectively. Finally, the second
harmonic amplitude is shown as dash-dot-dotted line. It's clear that $|A|$
is relatively small in the wide region $\Delta L$ and reaches the value of $%
|2B|$ only at local maximum. The increased width of $\Delta L$ (see Eqs.
(29),(49)) is provided by 
geometric attributes of RTO type structure.

\begin{figure}[th]
\begin{minipage}[h]{0.9\linewidth}
\center{\includegraphics[width=1\linewidth]{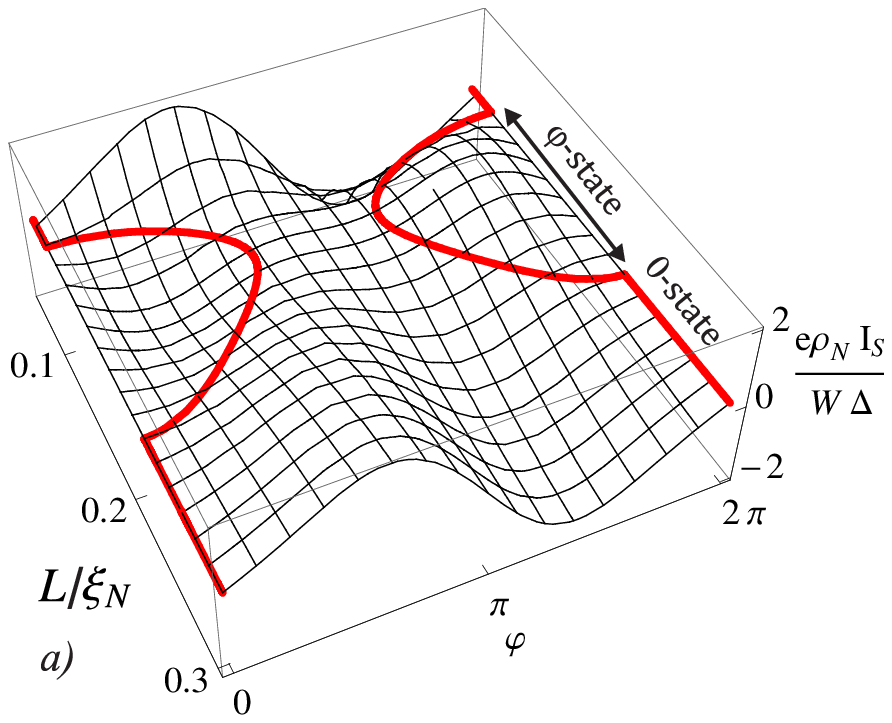}} \\
\end{minipage}
\vfill
\begin{minipage}[h]{0.9\linewidth}
\center{\includegraphics[width=1\linewidth]{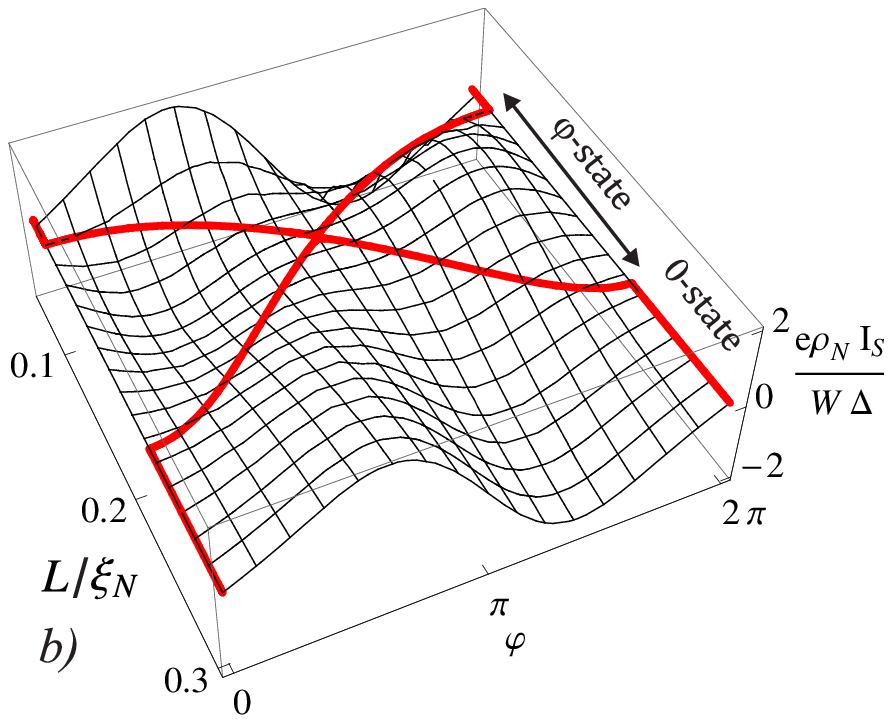}} \\
\end{minipage}
\vfill
\begin{minipage}[h]{0.9\linewidth}
\center{\includegraphics[width=1\linewidth]{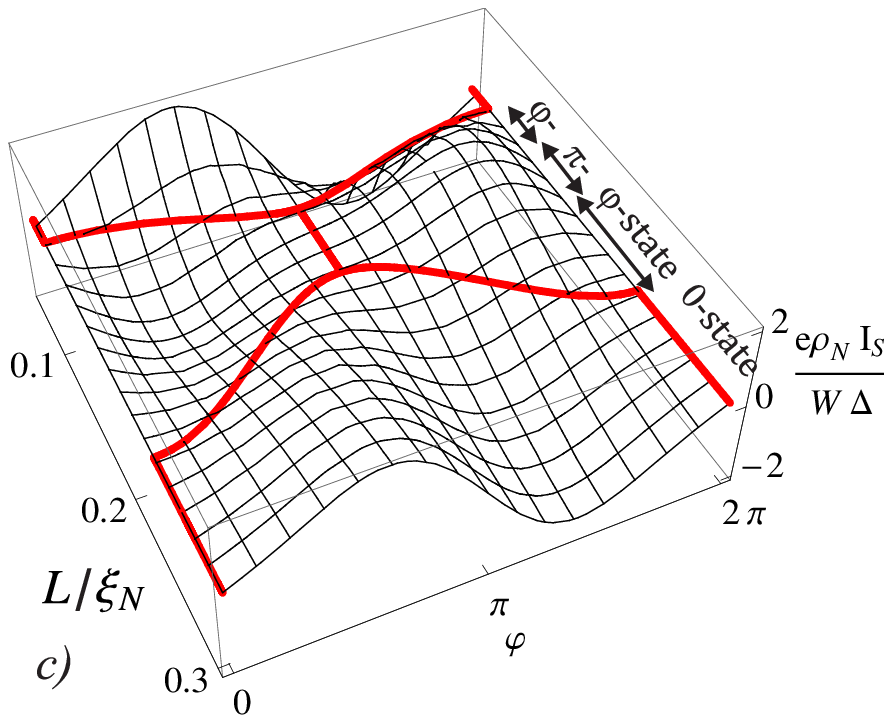}} \\
\end{minipage}
\caption{{}The full current $I_{S}$ versus Josephson phase $\protect\varphi $
and electrode spacing $L$ for RTO structure at $T\ll T_{C},$ $\protect\gamma %
_{BM}=0.64$ and at different F-layer thickness parameters a) $\protect%
\varepsilon = 0.137$, b) $\protect\varepsilon = 0.123$, c) $\protect%
\varepsilon = 0.111$. The lines mark the ground states phase $\protect%
\varphi _{g}.$}
\label{RTO_3d1}
\end{figure}

Let us illustrate the range of nontrivial ground phase $\varphi _{g}$
existence in the structure described in Fig.\ref{RTO_ideal}. The total
supercurrent $I_{S}$ is shown on Fig.\ref{RTO_3d1} as a function of
Josephson phase $\varphi $ and electrode spacing $L$. It means that each $L-$%
section of this 3D graph 
is CPR. Solid lines mark the ground state phases at each $L$. In the range
of small and large spacing $L$ 
ground phase is located at $\varphi _{g}=0$. However, in the $\Delta L $%
-interval CPR becomes significantly nonsinusoidal and demands ground phase $%
\varphi _{g}$ to split and go to $\pi $ from both sides; 
then $\pi-$state is realized at $\kappa L=\pi /2$. Clearly, for $\varepsilon
\gtrsim 0.123$ the value $\varphi _{g}=\pi $ can not be reached (see Fig.\ref%
{RTO_3d1}a), while in the case of $\varepsilon \lesssim 0.123$ the prolonged 
$\pi $-state region is formed (see Fig.\ref{RTO_3d1}c).

\section{Discussion \label{discuss}}

We have shown that stable $\varphi $-state can be realized in S-NF-S
structures with longitudinally oriented NF-bilayers (though $\varphi $-state
can not be achieved in conventional SNS and SFS structures). We have
discussed the conditions for realization of $\varphi $-state in ramp-type
S-NF-S and RTO-type SN-FN-NS geometries. %

Let us discuss most favorable conditions for for experimental realization of 
$\varphi $-junction. We suggest to use Copper as a normal film ($\xi
_{N}\approx 100\;nm$ and $\rho =5\ast 10^{-8}\;\Omega m$) and strongly
diluted ferromagnet like FePd or CuNi alloy $(\xi _{F}\approx
10\;nm,~H\approx 10T_{C})$ as the F-layer. We chose Nb $(T_{C}\approx 9 K)$
as a superconducting electrode material since it is commonly used in
superconducting circuits applications. We also propose to use sufficiently
thick normal layer, above the saturation threshold, when N-layer thickness
have almost no effect. 
%
%
After substitution of relevant values into (\ref{ineq4}) and (\ref{boun_kl3}%
) we arrived at a fairly broad geometrical margins, within which there is a
possibility for implementation of $\varphi $-junctions 
\begin{eqnarray}
d_{N} &\gtrsim &50\;nm,  \notag \\
60\;nm &\lesssim &d_{F}\lesssim 150\;nm, \\
7\;nm &\lesssim &L\lesssim 22\;nm.  \notag
\end{eqnarray}%
Finally, the last out-of-plane geometrical scale is set as $W=140~nm$. This
value maximizes current and conserves the scale of structure in a range of $%
100~nm$. The magnitude of critical supercurrent in the $\varphi $-state is
determined by the second harmonic amplitude $B$ 
\begin{equation}
I_{C}\sim B_{N}=\frac{2W\Delta }{e\rho _{N}\gamma _{BN}}\Upsilon _{B}\approx
1~mA.
\end{equation}%
The spreads of geometrical scales as well as the magnitude of critical
current are large enough to be realized experimentally.

By creating $\varphi $-state in a Josephson junction one can fix certain
value of ground phase $\varphi _{g}$. Temperature variation slightly shifts
the interval of relevant $0$-$\pi $ transition and permits one to tune the
desired ground state phase. Furthermore, sensitivity of the ground state to
an electron distribution function permits $\varphi $-junctions to be applied
as small-scale self-biasing one-photon detectors. Moreover, quantum
double-well potential is formed at the point of ground state splitting
providing necessary condition for quantum bits and quantum detectors. To
summarize, Josephson $\varphi $-junctions can be realized using up-to-date
technology and may become important basic element in superconducting
electronics.

\ 

\begin{acknowledgments}
We gratefully acknowledge V. V. Ryazanov helpful discussions. This work was
supported by the Russian Foundation for Basic Research (Grants
12-02-90010-Bel-a, 11-02-12065-ofi-m), Russian Ministry of Education and
Science, Dynasty Foundation and Dutch FOM.
\end{acknowledgments}

\appendix

\section{Ramp type junctions. Limit of small $L.$ \label{A}}




In the limit of small spacing between S electrodes 
\begin{equation}
L\ll \min \{\xi _{F},\xi _{N}\}  \label{AsmallL}
\end{equation}%
%
%
%
%
%
%
%
%
%
%
%
%
%
%
%
%
%
%
%
%
%
we can neglect nongradient terms in (\ref{UsadelFull}) 
\begin{equation}
\frac{\partial }{\partial x}\left( G_{F,N}^{2}\frac{\partial }{\partial x}%
R_{F,N}\right) +\frac{\partial }{\partial z}\left( G_{F,N}^{2}\frac{\partial 
}{\partial z}R_{F,N}\right) =0,  \label{ALapR}
\end{equation}%
\begin{equation}
\frac{\partial }{\partial x}\left( G_{F,N}^{2}\frac{\partial }{\partial x}%
U_{F,N}\right) +\frac{\partial }{\partial z}\left( G_{F,N}^{2}\frac{\partial 
}{\partial z}U_{F,N}\right) =0,  \label{ALapI}
\end{equation}%
and introduce four functions%
\begin{equation}
\Phi _{F}=R_{F}+iU_{F},\ \Phi _{N}=R_{N}+iU_{N},  \label{ARI}
\end{equation}%
where, $i,$ is imaginary unit, $R_{F}$ and $R_{N}$ are even function of
coordinate $x,$ while $U_{F}$ and $U_{N}$ are odd in $x.$ Due to the
symmetry at $x=0$%
\begin{equation}
\frac{\partial R_{F,N}}{\partial x}=0,\ U_{F,N}=0  \label{Acond0}
\end{equation}%
for any coordinate $z,$ and it is convenient to rewrite boundary conditions (%
\ref{Gr8}), (\ref{Gr9}) at $x=L/2$ in the form 
\begin{subequations}
\label{GrX}
\begin{align}
\gamma _{BN}\xi _{N}\frac{\partial R_{N}}{\partial x}& =\frac{G_{S}}{G_{N}}%
\left( \Delta \cos (\varphi /2)-R_{N}\right) ,  \label{AReLN} \\
\gamma _{BF}\xi _{F}\frac{\partial R_{F}}{\partial x}& =\frac{G_{S}}{G_{F}}%
\left( \frac{\widetilde{\omega }}{\omega }\Delta \cos (\varphi
/2)-R_{F}\right) ,  \label{AImLN}
\end{align}%
\end{subequations}
\begin{subequations}
\label{GrX}
\begin{align}
\gamma _{BN}\xi _{N}\frac{\partial U_{N}}{\partial x}& =\frac{G_{S}}{G_{N}}%
\left( \Delta \sin (\varphi /2)-U_{N}\right) ,  \label{AReLF} \\
\gamma _{BF}\xi _{F}\frac{\partial U_{F}}{\partial x}& =\frac{G_{S}}{G_{F}}%
\left( \frac{\widetilde{\omega }}{\omega }\Delta \sin (\varphi
/2)-U_{F}\right) .  \label{AImFL}
\end{align}%
At NF interface the boundary conditions transforms to: 
\end{subequations}
\begin{subequations}
\begin{align}
\gamma _{BFN}\xi _{F}\frac{\partial R_{F}}{\partial z}& =-\frac{G_{N}}{G_{F}}%
\left( R_{F}-\frac{\widetilde{\omega }}{\omega }R_{N}\right) ,  \label{ANFRF}
\\
\gamma _{BNF}\xi _{N}\frac{\partial R_{N}}{\partial z}& =\frac{G_{F}}{G_{N}}%
\left( R_{N}-\frac{\omega }{\widetilde{\omega }}R_{F}\right) ,  \label{ANFRN}
\end{align}%
\end{subequations}
\begin{subequations}
\begin{align}
\gamma _{BFN}\xi _{F}\frac{\partial U_{F}}{\partial z}& =-\frac{G_{N}}{G_{F}}%
\left( U_{F}-\frac{\widetilde{\omega }}{\omega }U_{N}\right) ,  \label{ANFIF}
\\
\gamma _{BNF}\xi _{N}\frac{\partial U_{N}}{\partial z}& =\frac{G_{F}}{G_{N}}%
\left( U_{N}-\frac{\omega }{\widetilde{\omega }}U_{F}\right) .  \label{ANFIN}
\end{align}%
From (\ref{Acond0}) and (\ref{AReLN}) - (\ref{AImFL}) it follows that for $%
\gamma _{BF}$ and $\gamma _{BN}$ within the interval 
\end{subequations}
\begin{equation}
\frac{L}{\xi _{N}}\ll \gamma _{BN}\ll \frac{\xi _{N}}{L},\ \frac{L}{\xi _{1}}%
\ll \gamma _{BF}\ll \frac{\xi _{1}}{L},  \label{AcondGam}
\end{equation}%
we can neglect $U_{N,F}$ in left hand side of (\ref{AReLF}), (\ref{AImFL}).
Moreover, in this approximation for any point inside the weak link region $%
R_{F,N}\gg U_{F,N}$ and the boundary problem (\ref{ALapR})-(\ref{ANFIN}) for
functions $R_{F}$ and $R_{N}$ can be solved resulting in 
\begin{equation}
R_{N}=\Delta \cos (\varphi /2),\ R_{F}=\frac{\widetilde{\omega }}{\omega }%
\Delta \cos (\varphi /2)  \label{ARNRF}
\end{equation}%
and 
\begin{equation}
G_{N}=G_{F}=\frac{\omega }{\sqrt{\omega ^{2}+\Delta ^{2}\cos ^{2}(\varphi /2)%
}}  \label{AGNGF}
\end{equation}%
Therefore under conditions (\ref{AcondGam}) both $G_{N}$ and $G_{F}$ are
independent on coordinate $x,z$ functions and equations for $U_{F,N}$
transform to Laplas equations, which have the solutions%
\begin{equation}
\begin{array}{c}
U_{N}=\frac{\Delta \sin (\varphi /2)}{\gamma _{BN}}\frac{G_{S}}{G_{N}}\frac{x%
}{\xi _{N}}+ \\ 
+\sum\limits_{n=1}^{\infty }a_{n}\sin \frac{\pi (2n+1)x}{L}\cosh \frac{\pi
(2n+1)\left( z-d_{N}-d_{F}\right) }{L},%
\end{array}
\label{AIN1}
\end{equation}%
\begin{equation}
\begin{array}{c}
U_{F}=\frac{\Delta \sin (\varphi /2)}{\gamma _{BF}}\frac{\widetilde{\omega }%
}{\omega }\frac{G_{S}}{G_{F}}\frac{x}{\xi _{F}}+ \\ 
+\frac{\widetilde{\omega }}{\omega }\sum\limits_{n=1}^{\infty }b_{n}\sin 
\frac{(2n+1)\pi x}{L}\cosh \frac{\pi (2n+1)z}{L}.%
\end{array}
\label{AIF1}
\end{equation}%
They automatically satisfy the boundary conditions at $z=0$ and $%
z=d_{N}+d_{F},$ as well as at $x=0$ and $x=L/2.$ To find the integration
constants $a_{n}$ and $b_{n}$ we have to substitute (\ref{AIN1}) and (\ref%
{AIF1}) into (\ref{ANFIF}), (\ref{ANFIN}) and get%
\begin{equation}
\begin{array}{c}
a_{n}=-\frac{\Delta \sin (\varphi /2)G_{S}\Theta \gamma _{BFN}\xi _{F}t_{n}}{%
G_{N}\beta \cosh \frac{\pi (2n+1)d_{N}}{L}},\ t_{n}=\tanh \frac{\pi
(2n+1)d_{N}}{L}, \\ 
b_{n}=\frac{\Delta \sin (\varphi /2)G_{S}\Theta \gamma _{BNF}\xi _{N}t_{f}}{%
G_{N}\beta \cosh \frac{\pi (2n+1)d_{F}}{L}},\ t_{f}=\tanh \frac{\pi
(2n+1)d_{F}}{L},%
\end{array}%
\end{equation}%
where%
\begin{equation*}
\beta =\left( \gamma _{BNF}\xi _{N}\frac{\pi (2n+1)}{L}t_{n}+1\right) \gamma
_{BFN}\xi _{F}t_{f}+\gamma _{BNF}\xi _{N}t_{n},
\end{equation*}%
and 
\begin{equation*}
\Theta =\left( \frac{1}{\gamma _{BN}\xi _{N}}-\frac{1}{\gamma _{BF}\xi _{F}}%
\right) \frac{4L}{\pi ^{2}}\frac{\left( -1\right) ^{n}}{(2n+1)^{2}}.
\end{equation*}%
Substitution of (\ref{AIN1}) and (\ref{AIF1}) into expression for the
supercurrent (\ref{Ic_gen}) gives that contributions to the supercurrent
across the junction proportional to $a_{n}$ and $b_{n}$ cancel each other
and $I_{S}(\varphi )$ equals to the sum 
\begin{equation*}
I_{S}(\varphi )=I_{N}(\varphi )+I_{F}(\varphi ),
\end{equation*}%
%
%
%
%
%
%
%
%
%
%
%
%
%
%
%
%
%
%
%
%
\begin{equation}
\frac{2eI_{N}(\varphi )}{\pi TWd_{N}}=\frac{1}{\gamma _{BN}\xi _{N}\rho _{N}}%
\sum_{\omega =-\infty }^{\infty }\frac{\Delta ^{2}G_{N}G_{S}\sin (\varphi )}{%
\omega ^{2}},  \label{AIsmallL}
\end{equation}%
\begin{equation}
\frac{2eI_{F}(\varphi )}{\pi TWd_{F}}=\frac{1}{\gamma _{BF}\xi _{F}\rho _{F}}%
\sum_{\omega =-\infty }^{\infty }\frac{\Delta ^{2}G_{F}G_{S}\sin (\varphi )}{%
\omega ^{2}}  \label{IF_small_}
\end{equation}%
of the currents, $I_{N}(\varphi ),$ and, $I_{F}(\varphi ),$ flowing
independently across F and N parts of the weak link. 
%

\section{Ramp type junctions. Limit of intermediate $L.$\label{B}}


For intermediate values of spacing between the S electrodes%
\begin{equation}
\xi _{1}\ll L\ll \xi _{N}.  \label{interL}
\end{equation}%
and suppression parameters at SN and SF interfaces belonging to the interval
(\ref{condgamma}) the boundary problem (\ref{UsadelFull})-(\ref{Gr9}) can be
also solved analytically for sufficiently large suppression parameter $%
\gamma _{BFN}.$ Under these restrictions in the first approximation we can
neglect the suppression of superconductivity in the N film due to proximity
with the F layer and use expressions (\ref{ARNRF}) and (\ref{AIF1}) with $%
a_{n}=0$ as the solution in the N part of the weak link.

To find $R_{F}$ and $U_{F}$ we have to solve the linear equations 
\begin{equation}
\xi _{F}^{2}\frac{\partial ^{2}}{\partial x^{2}}R_{F}+\xi _{F}^{2}\frac{%
\partial ^{2}}{\partial z^{2}}R_{F}-\widetilde{\Omega }R_{F}=0,
\label{linRF}
\end{equation}%
\begin{equation}
\xi _{F}^{2}\frac{\partial ^{2}}{\partial x^{2}}U_{F}+\xi _{F}^{2}\frac{%
\partial ^{2}}{\partial z^{2}}U_{F}-\widetilde{\Omega }U_{F}=0,
\label{linIF}
\end{equation}%
with the boundary conditions 
\begin{equation}
\gamma _{BF}\xi _{F}\frac{\partial R_{F}}{\partial x}=G_{S}\frac{\widetilde{%
\Omega }}{\Omega }\Delta \cos (\varphi /2),  \label{bcRFprom}
\end{equation}%
\begin{equation}
\gamma _{BF}\xi _{F}\frac{\partial U_{F}}{\partial x}=G_{S}\frac{\widetilde{%
\Omega }}{\Omega }\Delta \sin (\varphi /2),  \label{bcIFprom}
\end{equation}%
at $x=L/2,$ $0\leq z\leq d_{F}$ and 
\begin{equation}
\gamma _{BFN}\xi _{F}\frac{\partial R_{F}}{\partial z}=\frac{\widetilde{%
\Omega }}{\Omega }G_{N}R_{N},  \label{bcRFprom2}
\end{equation}%
\begin{equation}
\gamma _{BFN}\xi _{F}\frac{\partial U_{F}}{\partial z}=\frac{\widetilde{%
\Omega }}{\Omega }G_{N}U_{N},  \label{bcIFprom2}
\end{equation}%
at $z=d_{F},$ $0\leq x\leq L/2;\;$($\Omega =\omega /\pi T_{C},$ $\widetilde{%
\Omega }=\widetilde{\omega }sign(\omega )/\pi T_{C}).$ The boundary problem (%
\ref{linRF})-(\ref{bcIFprom2}) must be closed by the conditions (\ref{Gr5})
and (\ref{Acond0}) at free interface of the F film and at the line of
junction symmetry, respectively.

Spatial distribution of even in coordinate $x$ part of $\Phi _{F}(x,z)$ can
be found in the form of superposition of superconducting correlations
induced into F film from superconductors and from the N part of weak link%
\begin{equation}
\begin{array}{c}
R_{F}=\frac{\sqrt{\widetilde{\Omega }}G_{S}\Delta \cos \left( \varphi
/2\right) }{\Omega \gamma _{BF}}\frac{\cosh \left( \sqrt{\widetilde{\Omega }}%
\frac{x}{\xi _{F}}\right) }{\sinh \left( \sqrt{\widetilde{\Omega }}\frac{L}{%
2\xi _{F}}\right) }+ \\ 
+\frac{\sqrt{\widetilde{\Omega }}G_{N}\Delta \cos \left( \varphi /2\right) }{%
\Omega \gamma _{BFN}}\frac{\cosh \left( \sqrt{\widetilde{\Omega }}\frac{z}{%
\xi _{F}}\right) }{\sinh \left( \sqrt{\widetilde{\Omega }}\frac{d_{F}}{\xi
_{F}}\right) }.%
\end{array}
\label{RFSRL}
\end{equation}%
Solution for the odd part of $\Phi _{F}(x,z)$ consists of three terms%
\begin{equation}
\begin{array}{c}
U_{F}=\frac{\sqrt{\widetilde{\Omega }}G_{S}\Delta \sin \left( \varphi
/2\right) }{\Omega \gamma _{BN}\gamma _{BFN}}\frac{x\cosh \left( \sqrt{%
\widetilde{\Omega }}\frac{z}{\xi _{F}}\right) }{\xi _{N}\sinh \left( \sqrt{%
\widetilde{\Omega }}\frac{d_{F}}{\xi _{F}}\right) }- \\ 
-\frac{\widetilde{\Omega }^{3/2}G_{S}\Delta \sin \left( \varphi /2\right)
\xi _{F}^{2}}{\Omega \gamma _{BN}\xi _{N}\gamma _{BFN}d_{F}}%
\sum\limits_{n=-\infty }^{\infty }\frac{\left( -1\right) ^{n}\cos \left( 
\frac{\pi nz}{d_{F}}\right) \sinh \left( \kappa _{n}\frac{x}{\xi _{F}}%
\right) }{\kappa _{n}^{3}\cosh \left( \kappa _{n}\frac{L}{2\xi _{F}}\right) }%
+ \\ 
+\frac{\sqrt{\widetilde{\Omega }}G_{S}\Delta \sin \left( \varphi /2\right) }{%
\Omega \gamma _{BF}}\frac{\sinh \left( \sqrt{\widetilde{\Omega }}\frac{x}{%
\xi _{F}}\right) }{\cosh \left( \sqrt{\widetilde{\Omega }}\frac{L}{2\xi _{F}}%
\right) },%
\end{array}
\label{UFSRL}
\end{equation}%
where $\kappa _{n}^{2}=\widetilde{\Omega }+\left( \pi n\xi _{F}/d_{F}\right)
^{2}.$ The first two give the part of $U_{F}$ induced from the N film, while
the last has the well known for SFS junction form\cite{RevG}$^{,}$\cite{RevB}%
$^{,}$\cite{RevV}.

From (\ref{RFSRL}) and (\ref{UFSRL}) it follows that $R_{-\omega ,F}^{\ast
}=R_{\omega ,F}$ and $U_{-\omega ,F}^{\ast }=U_{\omega ,F}.$ Substitution of
(\ref{RFSRL}) and (\ref{UFSRL}) into expression for the supercurrent (\ref%
{Ic_gen}) gives that the $I_{S}(\varphi )$ dependence is consists of three
terms%
\begin{equation}
I_{S}(\varphi )=I_{N}(\varphi )+I_{F}(\varphi )+I_{FN}(\varphi ).
\label{IotFimiddle}
\end{equation}%
The first is the supercurrent across the N layer. In considered
approximation it coincides with the expression given by (\ref{AIsmallL}).
The second term in (\ref{IotFimiddle}) is the supercurrent across SFS double
barrier structure in the limit of small transparencies of SF interfaces\cite%
{Golubov1}$^{,}$\cite{DBbuz}%
\begin{equation}
\frac{2eI_{F}(\varphi )}{\pi TWd_{F}}=\frac{\Delta ^{2}\sin \left( \varphi
\right) }{\gamma _{BF}^{2}\xi _{F}\rho _{F}}\sum_{\omega =-\infty }^{\infty }%
\frac{G_{S}^{2}}{\omega ^{2}\sqrt{\widetilde{\Omega }}\sinh \left(
2q_{L}\right) }  \label{IotFiSFS}
\end{equation}%
and the last consists of two terms, $I_{FN}(\varphi )=I_{1}(\varphi
)+I_{2}(\varphi )$ having different $\varphi -$dependence%
\begin{equation}
\begin{array}{c}
\frac{2eI_{1}(\varphi )}{\pi TWd_{F}}=\frac{\Delta ^{2}\sin \left( \varphi
\right) }{\rho _{F}d_{F}}\frac{\xi _{F}}{\gamma _{BF}\gamma _{BFN}\gamma
_{BN}\xi _{N}}\sum\limits_{\omega =-\infty }^{\infty }\frac{G_{S}^{2}}{%
\widetilde{\Omega }^{2}\omega ^{2}}\Psi _{1}, \\ 
\Psi _{1}=\frac{\sqrt{\widetilde{\Omega }}}{\sinh (q_{L})}-\frac{2\widetilde{%
\Omega }}{\sinh \left( 2q_{L}\right) },%
\end{array}
\label{I1otFi}
\end{equation}%
\begin{equation}
\begin{array}{c}
\frac{2eI_{2}(\varphi )}{\pi TWd_{F}}=\frac{\Delta ^{2}\sin \left( \varphi
\right) }{\gamma _{BFN}\rho _{F}d_{F}}\sum\limits_{\omega =-\infty }^{\infty
}\frac{G_{N}G_{S}}{\omega ^{2}\widetilde{\Omega }^{2}}\left( \frac{1}{\gamma
_{BN}\gamma _{BFN}\xi _{N}}\Psi _{2}+\frac{\widetilde{\Omega }}{\gamma
_{BF}\cosh q_{L}}\right) , \\ 
\Psi _{2}=\frac{d_{F}\widetilde{\Omega }\left( 2q_{d}+\sinh \left(
2q_{d}\right) \right) }{4q_{d}\sinh ^{2}(q_{d})}-\frac{\widetilde{\Omega }%
\xi _{F}}{q_{d}\cosh \left( q_{L}\right) }-\sum\limits_{n=1}^{\infty }\frac{2%
\widetilde{\Omega }^{3}\xi _{F}}{q_{d}\kappa _{n}^{4}\cosh \left( \frac{%
L\kappa _{n}}{2\xi _{F}}\right) },%
\end{array}
\label{I2otFi}
\end{equation}%
where $q_{d}=d_{F}\sqrt{\widetilde{\Omega }}/\xi _{F},$ $q_{L}=L\sqrt{%
\widetilde{\Omega }}/2\xi _{F}.$ In real experimental situation 
\begin{equation}
\xi _{F}\ll \xi _{N},\ d_{F}\gg \xi _{F}.  \label{explim}
\end{equation}%
Under the conditions (\ref{explim}) some terms of $I_{FN}(\varphi )$ can be
neglected. Still existing expressions of it parts $I_{FN1}(\varphi )$ - $%
I_{FN3}(\varphi )$ simplify to 
\begin{equation}
\text{{\small $\frac{2eI_{FN1}(\varphi )}{\pi TWd_{F}}=\frac{\Delta ^{2}\sin
\left( \varphi \right) }{\gamma _{BF}\gamma _{BFN}\gamma _{BN}\rho _{F}d_{F}}%
\frac{\xi _{F}}{\xi _{N}}\sum_{\omega =-\infty }^{\infty }\frac{G_{S}^{2}}{%
\omega ^{2}\widetilde{\Omega }^{2}}\frac{\sqrt{\widetilde{\Omega }}}{\sinh
q_{L}}$,}}  \label{I1}
\end{equation}%
\begin{equation}
\text{{\small $\frac{2eI_{FN2}(\varphi )}{\pi TWd_{F}}$}}{\small =}\text{%
{\small $\frac{\Delta ^{2}\sin \left( \varphi \right) }{2\gamma _{BN}\gamma
_{BFN}^{2}\rho _{F}d_{F}}\sum_{\omega =-\infty }^{\infty }\frac{G_{N}G_{S}}{%
\omega ^{2}\widetilde{\Omega }^{3/2}}\frac{\xi _{F}}{\xi _{N}},$}}
\label{I2}
\end{equation}%
\begin{equation}
\text{{\small $\frac{2eI_{FN3}(\varphi )}{\pi TWd_{F}}$}}{\small =}\text{%
{\small $\frac{\Delta ^{2}\sin \left( \varphi \right) }{\gamma _{BFN}\gamma
_{BF}\rho _{F}d_{F}}\sum_{\omega =-\infty }^{\infty }\frac{G_{N}G_{S}}{%
\omega ^{2}\widetilde{\Omega }}$}}\frac{1}{\cosh q_{L}},  \label{I3}
\end{equation}%
.


\section{Overlap SN-N-NS junctions.\label{C}}

To calculate critical current of SN-N-NS junctions we consider the most
practical case of thin N film 
\begin{equation}
d_{N}\ll \xi _{N}  \label{condThinN}
\end{equation}%
and sufficiently large $\gamma _{BFN}$ providing the absence of suppression
of superconductivity in N film due to proximity with F layer. We will also
assume that electrode spacing $L$ is also small%
\begin{equation}
L\ll \xi _{N},  \label{smL}
\end{equation}%
in order to have nonsinusoidal CPR.

Condition (\ref{condThinN}) permits to perform averaging of Usadel equations
in $z-$direction in N film, as it was described in detail in \cite%
{Karminskaya}, and reduce the problem to the solution of one dimensional
equations for $\Phi _{N}=R_{N}+iU_{N}$. The real part of $\Phi _{N}$ is the
solution of the boundary problem%
\begin{equation}
\text{{\small $\frac{\xi _{N}^{2}\gamma _{BM}}{G_{N}\left( \Omega \gamma
_{BM}+G_{S}\right) }\frac{\partial }{\partial x}\left( G_{N}^{2}\frac{%
\partial R_{N}}{\partial x}\right) -R_{N}=-r\Delta $}$\cos $}\frac{\text{%
{\small $\varphi $}}}{{\small 2}}{\small ,~}\text{{\small $\frac{L}{2}\leq
x\leq \infty ,$}}  \label{RsmL1}
\end{equation}%
\begin{equation}
\frac{\xi _{N}^{2}}{\Omega G_{N}}\frac{\partial }{\partial x}\left( G_{N}^{2}%
\frac{\partial R_{N}}{\partial x}\right) =0,\ 0\leq x\leq \frac{L}{2},
\label{RsmL2}
\end{equation}%
\begin{equation}
\frac{\partial R_{N}}{\partial x}=0,\ \ x=0,\ \ x\rightarrow \infty ,
\label{BcRsmL1}
\end{equation}%
where $r=G_{S}/\left( \Omega \gamma _{BM}+G_{S}\right) ,$ $\gamma
_{BM}=\gamma _{BN}d_{N}/\xi _{N},~\delta =\Delta /\pi T_{C}.$

From (\ref{RsmL2}), (\ref{BcRsmL1}) it follows that at $0\leq x\leq L/2$
functions $R_{N}$ are independent on $x$ constants resulting in 
\begin{equation}
\frac{\partial R_{N}}{\partial x}\left( \frac{L}{2}\right) =0.  \label{Bcqq}
\end{equation}%
The arising boundary problem (\ref{RsmL1}), (\ref{BcRsmL1}), (\ref{Bcqq}) is
also satisfied by independent on $x$ constants leading to 
\begin{equation}
R_{N}=r\Delta \cos (\varphi /2),\ 0\leq x<\infty .  \label{Re}
\end{equation}%
Introducing now new functions, $\theta $ 
\begin{equation}
U_{N}=\mu \tan \theta ,\ G_{N}=\frac{\Omega }{\mu }\cos \theta ,
\label{tetap}
\end{equation}%
where $\mu =\sqrt{\Omega ^{2}+r^{2}\delta ^{2}\cos ^{2}(\varphi /2)},$ we
get 
\begin{equation}
\lambda ^{2}\frac{\partial ^{2}}{\partial x^{2}}\theta -\sin \left( \theta
-\phi \right) =0,\ \ \frac{L}{2}\leq x<\infty ,  \label{EqForTeta1}
\end{equation}%
\begin{equation}
\frac{\xi _{N}^{2}}{\cos \theta }\frac{\partial ^{2}}{\partial x^{2}}\theta
=0,\ 0\leq x\leq \frac{L}{2},  \label{EqForTeta2}
\end{equation}%
\begin{equation}
\theta (0)=0,\ \frac{\partial \theta }{\partial x}=0,\ \ \ x\rightarrow
\infty ,  \label{BCForTeta}
\end{equation}%
where 
\begin{equation}
\lambda =\xi _{N}\sqrt{\frac{\Omega \gamma _{BM}}{\left( \Omega \gamma
_{BM}+G_{S}\right) \sqrt{\Omega ^{2}+r^{2}\delta ^{2}}}},  \label{dzita}
\end{equation}%
\begin{equation}
\tan \phi =\frac{r\sin (\varphi /2)}{\mu }.  \label{tanFi}
\end{equation}%
Solution of Eq. (\ref{EqForTeta2}) can be easily found 
\begin{equation}
\theta (x)=\frac{2x}{L}\theta (\frac{L}{2}),\ 0\leq x\leq \frac{L}{2}.
\label{solI0L}
\end{equation}%
Solution of Eq. (\ref{EqForTeta1}) can be simplified due to existence of the
first integral%
\begin{equation}
\frac{\lambda ^{2}}{2}\left( \frac{\partial }{\partial x}\theta \right)
^{2}+\cos \left( \theta -\phi \right) =1.\   \label{FirstInt}
\end{equation}%
The constant of integration in the right hand side of (\ref{FirstInt}) have
been found from the boundary condition (\ref{BCForTeta}), which demands $%
\theta \rightarrow \phi $ then $x\rightarrow \infty .$ Further integration
in (\ref{FirstInt}) for \ $L/2\leq x<\infty $ gives 
\begin{equation}
\theta =\phi +4\arctan \left( C_{2}\exp \left( -\frac{x-L/2}{\lambda }%
\right) \right) ,  \label{SolTetaLx}
\end{equation}%
where $C_{2}$ is integration constant, which should be determined from the
matching conditions at $x=L/2.$ For $C_{2}$ they give 
\begin{equation}
\left( \phi +4\arctan \left( C_{2}\right) \right) =-\frac{2C_{2}}{1+C_{2}^{2}%
}\frac{L}{\lambda }.  \label{EqForC2}
\end{equation}%
Assuming additionally that $\gamma _{BM}$ is not too small, namely that $%
L\ll \xi _{N}\min \left( 1,\sqrt{\gamma _{BM}}\right) ,$ from (\ref{EqForC2}%
) it is easy to get%
\begin{equation}
C_{2}=-\tan \left( \frac{\phi }{4}-\frac{L}{4\lambda }\sin \frac{\phi }{2}%
\right) ,  \label{C2}
\end{equation}%
resulting in 
\begin{equation}
\theta (x)=\frac{2x}{\lambda }\sin \frac{\phi }{2},\ 0\leq x\leq \frac{L}{2}.
\label{solTeta1}
\end{equation}%
From (\ref{solTeta1}) it follows that in weak link region $\left\vert
x\right\vert \leq L/2$ 
\begin{equation}
U_{N}=\frac{2x}{\lambda }\mu \sin \frac{\phi }{2},\ G_{N}=\frac{\Omega }{\mu 
},  \label{SolINGN}
\end{equation}%
while under the S electrode, $L/2\leq x<\infty $%
\begin{equation}
\begin{array}{c}
U_{N}=\mu \tan \left( \phi -4\arctan \left( u\right) \right) , \\ 
u=\tan \left( \frac{\phi }{4}-\frac{L}{4\lambda }\sin \frac{\phi }{2}\right)
\exp \left( -\frac{x-L/2}{\lambda }\right) .\ 
\end{array}%
\   \label{solUNunderS}
\end{equation}

%
%
Substitution of (\ref{Re}), (\ref{SolINGN}) into expression (\ref{Ic_gen})
for the supercurrent in the N channel results in 
\begin{equation}
\frac{2eI_{N}(\varphi )}{\pi TWd_{N}}=\frac{2}{\rho _{N}\xi _{N}\sqrt{\gamma
_{BM}}}\sum_{\omega =-\infty }^{\infty }\frac{r^{2}\delta ^{2}\sin \varphi 
\sqrt{\left( \Omega \gamma _{BM}+G_{S}\right) }}{\sqrt{2\Omega \mu
^{2}\left( \sqrt{\Omega ^{2}+r^{2}\delta ^{2}}+\mu \right) }}.  \label{RTOIN}
\end{equation}

\section{Solution in Ferromagnet Layer of RTO junction. \label{D}}

Spatial distribution of even and odd in coordinate $x$ parts of $\Phi
_{F}(x,z)$ can be found in the form of superposition of superconducting
correlations induced into F film from superconductors and from the N part of
weak link. It has the same form as in (\ref{RFSRL}) and (\ref{UFSRL})

\begin{equation}
\begin{array}{c}
R_{F}=\frac{\sqrt{\widetilde{\Omega }}G_{S}\Delta \cos \left( \varphi
/2\right) }{\Omega \gamma _{BF}}\frac{\cosh \left( \sqrt{\widetilde{\Omega }}%
\frac{x}{\xi _{F}}\right) }{\sinh \left( \sqrt{\widetilde{\Omega }}\frac{L}{%
2\xi _{F}}\right) }+ \\ 
+\frac{\sqrt{\widetilde{\Omega }}G_{N}R_{N}}{\Omega \gamma _{BFN}}\frac{%
\cosh \left( \sqrt{\widetilde{\Omega }}\frac{z}{\xi _{F}}\right) }{\sinh
\left( \sqrt{\widetilde{\Omega }}\frac{d_{F}}{\xi _{F}}\right) },%
\end{array}
\label{RFRTO}
\end{equation}%

\begin{equation}
\begin{array}{c}
U_{F}=\frac{\sqrt{\widetilde{\Omega }}G_{N}U_{N}}{\Omega \gamma _{BFN}}\frac{%
\cosh \left( \sqrt{\widetilde{\Omega }}\frac{z}{\xi _{F}}\right) }{\sinh
\left( \sqrt{\widetilde{\Omega }}\frac{d_{F}}{\xi _{F}}\right) }- \\ 
-\frac{\widetilde{\Omega }^{3/2}\xi _{F}^{2}G_{N}(U_{N}/x)}{\Omega \gamma
_{BFN}d_{F}}\sum\limits_{n=-\infty }^{\infty }\frac{\left( -1\right)
^{n}\cos \left( \frac{\pi nz}{d_{F}}\right) \sinh \left( \kappa _{n}\frac{x}{%
\xi _{F}}\right) }{\kappa _{n}^{3}\cosh \left( \kappa _{n}\frac{L}{2\xi _{F}}%
\right) }+ \\ 
+\frac{\sqrt{\widetilde{\Omega }}G_{S}\delta \sin \left( \varphi /2\right) }{%
\Omega \gamma _{BF}}\frac{\sinh \left( \sqrt{\widetilde{\Omega }}\frac{x}{%
\xi _{F}}\right) }{\cosh \left( \sqrt{\widetilde{\Omega }}\frac{L}{2\xi _{F}}%
\right) },%
\end{array}
\label{UFRTO}
\end{equation}%
with the functions $R_{N},$ $G_{N},$ and $U_{N}$ defined by equations
followed from the solution of the boundary problem in the N layer described
in Appendix \ref{C}.

\begin{equation}
R_{N}=r\Delta \cos (\varphi /2),\ \ G_{N}=\frac{\Omega }{\sqrt{\Omega
^{2}+r^{2}\delta ^{2}\cos ^{2}(\varphi /2)}},  \label{RNGN}
\end{equation}%
\begin{equation}
\begin{array}{c}
U_{N}=\alpha \Delta \sin (\varphi /2)\frac{G_{S}}{G_{N}}\frac{x}{\xi _{N}},
\\ 
\alpha =\frac{2\sqrt{\Omega ^{2}+\delta ^{2}}}{\sqrt{2\left( \sqrt{\Omega
^{2}+r^{2}\delta ^{2}}+\mu \right) }}\frac{r}{\sqrt{1-r}}.%
\end{array}
\label{UNRTO1}
\end{equation}%
Substitution of (\ref{RFRTO})-(\ref{UNRTO1}) into expression (\ref{Ic_gen})
gives that supercurrent \ across F layer in RTO junction consists of the sum
of $I_{F}(\varphi )$ and $I_{FN}(\varphi )$, where $I_{F}(\varphi )$ is the
current through one dimensional double barrier SFS structure defined by Eq. (%
\ref{IotFiSFS}), while $I_{FN}(\varphi )=I_{1}(\varphi )+I_{2}(\varphi )$
has the form \ 
\begin{equation}
\begin{array}{c}
\frac{2eI_{1}(\varphi )}{\pi TWd_{F}}=\frac{\Delta ^{2}\sin \left( \varphi
\right) }{\rho _{F}d_{F}}\frac{\xi _{F}}{\gamma _{BF}\gamma _{BFN}\xi _{N}}%
\sum\limits_{\omega =-\infty }^{\infty }\frac{\alpha G_{S}^{2}}{\widetilde{%
\Omega }^{2}\omega ^{2}}\Psi _{1}, \\ 
\Psi _{1}=\frac{\sqrt{\widetilde{\Omega }}}{\sinh (q_{L})}-\frac{2\widetilde{%
\Omega }}{\sinh \left( 2q_{L}\right) },%
\end{array}
\label{I1RTO}
\end{equation}%
\begin{equation}
\begin{array}{c}
\frac{2eI_{2}(\varphi )}{\pi TWd_{F}}=\frac{\Delta ^{2}\sin \left( \varphi
\right) }{\rho _{F}d_{F}}\frac{1}{\gamma _{BFN}}\sum\limits_{\omega =-\infty
}^{\infty }\frac{rG_{N}G_{S}}{\omega ^{2}\widetilde{\Omega }^{2}}\left( 
\frac{\alpha }{\gamma _{BFN}\xi _{N}}{\small \Psi }_{2}{\small +}\frac{%
\widetilde{\Omega }}{\gamma _{BF}\cosh q_{L}}\right) , \\ 
{\small \Psi }_{2}{\small =}\frac{d_{F}\widetilde{\Omega }\left(
2q_{d}+\sinh \left( 2q_{d}\right) \right) }{4q_{d}\sinh ^{2}(q_{d})}-\frac{%
\widetilde{\Omega }\xi _{F}}{q_{d}\cosh \left( q_{L}\right) }%
-\sum\limits_{n=1}^{\infty }\frac{2\widetilde{\Omega }^{3}\xi _{F}}{%
q_{d}\kappa _{n}^{4}\cosh \left( \frac{L\kappa _{n}}{2\xi _{F}}\right) }.%
\end{array}
\label{I2RTO}
\end{equation}%
Application of conditions (\ref{explim}) allows to neglect some terms in $%
I_{FN}(\varphi )=I_{FN1}(\varphi )+I_{FN2}(\varphi )+I_{FN3}(\varphi )$ and
to simplify remaining terms, leading to the following expressions:%
\begin{equation}
\text{{\small $\frac{2eI_{FN1}(\varphi )}{\pi TWd_{F}}=\frac{\Delta ^{2}\sin
\left( \varphi \right) }{\gamma _{BF}\gamma _{BFN}\rho _{F}d_{F}}\frac{\xi
_{F}}{\xi _{N}}\sum_{\omega =-\infty }^{\infty }\frac{\alpha G_{S}^{2}}{%
\omega ^{2}\widetilde{\Omega }^{2}}\frac{\sqrt{\widetilde{\Omega }}}{\sinh
q_{L}}$,}}  \label{A_FNimp4}
\end{equation}%
\begin{equation}
\text{{\small $\frac{2eI_{FN2}(\varphi )}{\pi TWd_{F}}$=$\frac{\Delta
^{2}\sin \left( \varphi \right) }{2\gamma _{BFN}^{2}\rho _{F}d_{F}}%
\sum_{\omega =-\infty }^{\infty }\frac{r\alpha G_{N}G_{S}}{\omega ^{2}%
\widetilde{\Omega }^{3/2}}\frac{\xi _{F}}{\xi _{N}},$}}  \label{A_FNimp5}
\end{equation}%
\begin{equation}
\text{{\small $\frac{2eI_{FN3}(\varphi )}{\pi TWd_{F}}$=$\frac{\Delta
^{2}\sin \left( \varphi \right) }{\gamma _{BFN}\gamma _{BF}\rho _{F}d_{F}}%
\sum_{\omega =-\infty }^{\infty }\frac{rG_{N}G_{S}}{\omega ^{2}\widetilde{%
\Omega }}$}}\frac{1}{\cosh q_{L}}.  \label{A_FNimp6}
\end{equation}


\begin{thebibliography}{99}
\bibitem{Likharev} K.K. Likharev, Rev. Mod. Phys. \textbf{51}, 101 (1979).

\bibitem{RevG} A.~A.~Golubov, M.~Yu.~Kupriyanov, E.~Il'ichev, Rev. Mod.
Phys. \textbf{76}, 411 (2004).

\bibitem{RevB} A.~I.~Buzdin, Rev.\ Mod.\ Phys.\ \textbf{77}, 935 (2005).

\bibitem{RevV} F.~S.~Bergeret, A.~F.~Volkov, K.~B.~Efetov, Rev.\ Mod.\
Phys.\ \textbf{77}, 1321 (2005).

\bibitem{ryazanov2001} V. V. Ryazanov, V. A. Oboznov, A. Yu. Rusanov, A. V.
Veretennikov, A. A. Golubov, and J. Aarts, Phys. Rev. Lett. \textbf{86},
2427 (2001).

\bibitem{Ryazanov2004} S. M. Frolov, D. J. Van Harlingen, V. A. Oboznov, V.
V. Bolginov, and V. V. Ryazanov, Phys. Rev. B \textbf{70}, 144505 (2004);

\bibitem{Kontos} T. Kontos, M. Aprili, J. Lesueur, F. Genet, B. Stephanidis,
and R. Boursier, Phys. Rev. Lett. \textbf{89}, 137007 (2002).

\bibitem{sellier} H. Sellier, C. Baraduc, F. Lefloch, and R. Calemczuck,
Phys. Rev. B \textbf{68}, 054531 (2003).

\bibitem{Blum} Y. Blum, A. Tsukernik, M. Karpovski, and A. Palevski, Phys.
Rev. B \textbf{70}, 214501 (2004).

\bibitem{Surgers} C. Surgers, T. Hoss, C. Schonenberger, C. Strunk, J. Magn.
Magn. Mater. \textbf{240}, 598 (2002).

\bibitem{Bell} C. Bell, R. Loloee, G. Burnell, and M. G. Blamire Phys. Rev.
B \textbf{71}, 180501 (R) (2005).

\bibitem{Ryazanov2006} S. M. Frolov, D. J. Van Harlingen, V. V. Bolginov, V.
A. Oboznov, and V. V. Ryazanov, Phys. Rev. B \textbf{74}, 020503 (2006).

\bibitem{Ryazanov2006a} V. A. Oboznov, V. V. Bol'ginov, A. K. Feofanov, V.
V. Ryazanov, and A. I. Buzdin, Phys. Rev. Lett. \textbf{96}, 197003 (2006).

\bibitem{Shelukhin} V. Shelukhin, A. Tsukernik, M. Karpovski, Y. Blum, K. B.
Efetov, A. F. Volkov, T. Champel, M. Eschrig, T. Lofwander, G. Schon, and A.
Palevski, Physical Review B \textbf{73}, 174506 (2006).

\bibitem{Weides} M. Weides, K. Tillmann, and H. Kohlstedt, Physica C \textbf{%
437-438}, 349 (2006).

\bibitem{Weides1} M. Weides, M. Kemmler, H. Kohlstedt, A. Buzdin, E.
Goldobin, D. Koelle, R. Kleiner, Appl. Phys. Lett. \textbf{89}, 122511
(2006).

\bibitem{Weides2} M. Weides, M. Kemmler, H. Kohlstedt, R. Waser, D. Koelle,
R. Kleiner, and E. Goldobin Physical Review Letters \textbf{97} 247001
(2006).

\bibitem{Weides3} J. Pfeiffer, M. Kemmler, D. Koelle, R. Kleiner, E.
Goldobin, M. Weides, A. K. Feofanov, J. Lisenfeld, and A. V. Ustinov,
Physical Review B \textbf{77}, 214506 (2008).

\bibitem{Sellier} H. Sellier, C. Baraduc, F. Lefloch, and R. Calemczuck,
Phys. Rev. Lett. \textbf{92}, 257005 (2004).

\bibitem{Born} F. Born, M. Siegel, E. K. Hollmann, H. Braak, A. A. Golubov,
D. Yu. Gusakova, and M. Yu. Kupriyanov, Phys. Rev. B. \textbf{74}, 140501
(2006).

\bibitem{Blamire} J. W. A. Robinson, S. Piano, G. Burnell, C. Bell, and M.
G. Blamire, Phys. Rev. Lett. \textbf{97}, 177003 (2006).

\bibitem{Blamire2} S. Piano, J. W.A. Robinson, G. Burnell, M. G. Blamire The
European Physical Journal B \textbf{58}, 123 (2007).

\bibitem{Blamire3} J. W. Robinson, S. Piano, G. Burnell, C. Bell, and M. G.
Blamire Physical Review B \textbf{76}, 094522 (2007).

\bibitem{Keizer} R.~S.~Keizer, S.~T.~B.~Goennenwein, T.~M.~Klapwijk,
G.~Miao, G.~Xiao, A.~Gupta, Nature \textbf{439}, 825 (2006).

\bibitem{BirgeLR} T.~S.~Khaire, M.~A.~Khasawneh, W.~P.~Pratt, Jr., and
N.~O.~Birge, Phys. Rev. Lett. \textbf{104}, 137002 (2010).

\bibitem{RobinsonLR} J. W. A. Robinson, J. D. S. Witt, and M. G. Blamire,
Science \textbf{329}, 59 (2010).

\bibitem{WangLR} J.Wang, M. Singh, M. Tian, N. Kumar, B. Liu, C. Shi, J. K.
Jain, N. Samarth, T. E. Mallouk and M. H. W. Chan, Nat. Phys. \textbf{6},
389 (2010).

\bibitem{AartsLR} M. S. Anwar, F. Czeschka, M. Hesselberth, M. Porcu, and J.
Aarts, Phys. Rev. B \textbf{82}, 100501(R) (2010).

\bibitem{AnwarLR} M. S. Anwar, M. Veldhorst, A. Brinkman, and J. Aarts,
Appl. Phys.Lett. \textbf{100}, 052602 (2012).

\bibitem{Rogalla} T. Ortlepp, Ariando, O. Mielke, C. J. M. Verwijs, K. F. K.
Foo, H. Rogalla, F. H. Uhlmann, and H. Hilgenkamp, Science \textbf{312},
1495 (2006).

\bibitem{Feofanov} A.K. Feofanov, V.A. Oboznov, V.V. Bol'ginov, \emph{et. al}%
., Nature Physics \textbf{6}, 593 (2010).

\bibitem{Ustinov} A. V. Ustinov and V. K. Kaplunenko, J. Appl. Phys. \textbf{%
94}, 5405 (2003).

\bibitem{Mints} R.G. Mints, Phys. Rev. B \textbf{57}, R3221 (1998).

\bibitem{Buzdin} A. Buzdin and A. E. Koshelev, Phys. Rev. B \textbf{67},
220504(R) (2003).

\bibitem{Pugach} N. G. Pugach, E. Goldobin, R. Kleiner, and D. Koelle, Phys.
Rev. B \textbf{81}, 104513 (2010).

\bibitem{SPIE} M.Yu. Kupriyanov, A.A. Golubov, M. Siegel, Proceedings of the
SPIE, \textbf{6260}, 62600S-1 (2006).

\bibitem{Gold} E. Goldobin, D. Koelle, R. Kleiner, and R.G. Mints, Phys.
Rev. Lett, \textbf{107}, 227001 (2011); H. Sickinger, A. Lipman, M Weides,
R.G. Mints, H. Kohlstedt, D. Koelle, R. Kleiner, and E. Goldobin, arXiv
1207.3013.

\bibitem{Gold2007} E. Goldobin, D. Koelle, R. Kleiner, and A. Buzdin, Phys.
Rev. B, \textbf{76}, 224523 (2007).

\bibitem{Klenov} N. V. Klenov, N. G.Pugach, A. V.Sharafiev, S. V.Bakurskiy,
V. K.Kornev, Physics of Solid State, \textbf{52}, 2246 (2010).

\bibitem{Vasenko} A. S. Vasenko, A. A. Golubov, M. Yu. Kupriyanov, and M.
Weides, Phys.Rev.B, \textbf{77} 134507 (2008).

\bibitem{Vasenko1} A. S. Vasenko, S. Kawabata, A. A. Golubov, M. Yu.
Kupriyanov, C. Lacroix, F. W. J. Hekking, Phys. Rev. B \textbf{84} 024524
(2011).

\bibitem{Buzdin1} F. Konschelle, J. Cayssol, A.I. Buzdin, Phys. Rev. B 
\textbf{78}, 134505 (2008).

\bibitem{Vinokur} M. Houzet, V. Vinokur, and F. Pistolesi, PRB \textbf{72},
220506 (2005)


\bibitem{Karminskaya} T. Yu. Karminskaya and M. Yu. Kupriyanov, Pis'ma Zh.
Eksp. Teor. Fiz. \textbf{85}, 343 (2007) [JETP Lett. \textbf{85}, 286
(2007)].

\bibitem{Karminskaya1} T. Yu. Karminskaya and M. Yu. Kupriyanov, Pis'ma Zh.
Eksp. Teor. Fiz. \textbf{85}, 343 (2007) [JETP Lett. \textbf{86}, 61 (2007)].

\bibitem{Karminskaya2} T. Yu. Karminskaya, M. Yu. Kupriyanov, and A. A.
Golubov, Pis'ma Zh.Eksp. Teor. Fiz. \textbf{87,} 657 (2008) [JETP Lett., 
\textbf{87}, 570 (2008)].

\bibitem{Karminskaya3} Karminskaya T. Yu., Golubov A. A., Kupriyanov M. Yu.,
Sidorenko A. S., Phys.Rev.B \textbf{79}, 214509 (2009).

\bibitem{Karminskaya4} T. Yu. Karminskaya, A. A. Golubov, M. Yu. Kupriyanov,
and A. S. Sidorenko Phys. Rev. B \textbf{81}, 214518 (2010).

\bibitem{Bakurskiy1} S. V. Bakurskiy, N. V. Klenov, T. Yu. Karminskaya, M.
Yu. Kupriyanov and V. K. Kornev, Solid State Phenomena, \textbf{190}, 401
(2012).

\bibitem{Bergeret} F. S. Bergeret, A. F. Volkov, and K. B. Efetov, Phys.
Rev. Lett. \textbf{86}, 3140 (2001).

\bibitem{Fominov} Ya. V. Fominov, N. M. Chtchelkatchev, and A. A. Golubov,
Phys. Rev. B \textbf{66}, 014507 (2002).

\bibitem{Furusaki} A. Furusaki and M. Tsukada, Solid State Commun. \textbf{78%
}, 299 (1991).

\bibitem{Furusaki1} A. Furusaki and M. Tsukada, Phys. Rev. B \textbf{43}, 10
164 (1991).

\bibitem{Ivanov} Z. G.Ivanov, M. Yu. Kupriyanov, K. K. Likharev, S. V.
Meriakri, and O. V. Snigirev, Fiz. Nizk. Temp. \textbf{7}, 560 (1981). [Sov.
J. Low Temp. Phys. \textbf{7}, 274 (1981)]

\bibitem{Zubkov} A. A.Zubkov, and M. Yu. Kupriyanov, Fiz. Nizk. Temp. 
\textbf{9}, 548 (1983) \ [Sov. J. Low Temp. Phys. \textbf{9}, 279 (1983)].

\bibitem{KuperD} M. Yu. Kupriyanov, Pis'ma Zh. Eksp. Teor. Fiz. \ \textbf{56}%
, 414 (1992). [JETP Lett. \textbf{56}, 399 (1992)].

\bibitem{Beasley} Demler, E. A., Arnold G. B., and Beasley M. R., Phys. Rev.
B \textbf{55}, 15174 (1997).

\bibitem{Usadel} K.~D.~Usadel, Phys.\ Rev.\ Lett.\ \textbf{25}, 507 (1970).

\bibitem{KL} M.~Yu.~Kuprianov and V.~F.~Lukichev, Sov.\ Phys.\ JETP \textbf{%
67}, 1163 (1988) [Zh.\ Eksp.\ Teor.\ Fiz.\ \textbf{94}, 139 (1988)].

\bibitem{Chtchelkatchev} N. M. Chtchelkatchev and I. S. Burmistrov, Phys.
Rev. B \textbf{68}, 140501(R) (2003).

\bibitem{Burmistrov} I. S. Burmistrov and N. M. Chtchelkatchev, Phys. Rev. B 
\textbf{72}, 144520 (2005).

\bibitem{Champel1} T. Champel and M. Eschrig, PRB \textbf{72}, 054523 (2005).

\bibitem{HouzetBuz} M. Houzet and A. I. Buzdin, Phys. Rev. B \textbf{74},
214507 (2006).

\bibitem{Maleki} M. A. Maleki and M. Zareyan Physical Review B \textbf{74},
144512 (2006).

\bibitem{FominovVE} Y. V. Fominov, A. F. Volkov, and K. B. Efetov, Phys.
Rev. B \textbf{75}, 104509 (2007).

\bibitem{Anishchanka} A. F. Volkov and A. Anishchanka, Phys. Rev. B \textbf{%
71}, 024501 (2005).

\bibitem{Volkov2008} A.F. Volkov, K.B. Efetov Phys Rev B \textbf{78}, 024519
(2008).

\bibitem{Crouzy} B. Crouzy, S. Tollis, D. A. Ivanov, Phys. Rev. B \textbf{76}%
, 134502 (2007).

\bibitem{Golubov1} A.~A.~Golubov, M.~Yu.~Kupriyanov, and Ya.~V.~Fominov,
JETP Lett. \textbf{75}, 709 (2002) [Pisma v ZhETF \textbf{75}, 588 (2002)].

\bibitem{DBbuz} A. Buzdin, JETP Lett. \textbf{78}, 1073 (2003) [Pisma v
ZhETF \textbf{78}, 1073 (2003)].

\bibitem{KLO} M. Yu.Kupriyanov, V. F. Lukichev, A. A. Orlikovskii,
Mikroelektronika \textbf{15}, 328 (1986) [Soviet Microelectronics \textbf{15}%
, 185 (1986)].


\end{thebibliography}
\end{document}